\documentclass[10pt]{article}
\usepackage[]{graphicx}
\usepackage[]{subfigure}
\usepackage{epsfig}
\usepackage{bm}
\usepackage{amsfonts}
\usepackage{amsmath,amssymb}
\usepackage{dcolumn}
\PassOptionsToPackage{linktocpage}{hyperref}
\usepackage[draft=false]{hyperref}
\usepackage{threeparttable}
\usepackage[draft=false]{hyperref}
\usepackage{threeparttable}
\newcommand{\sfrac}[2]{{\textstyle{#1\over#2}}}
\def\case#1/#2{\textstyle\frac{#1}{#2}}
\newcommand{\ds}{\displaystyle}

\def\bea{\begin{eqnarray}}
\def\eea{\end{eqnarray}}

\newcommand{\vphi}{\varphi}

\newcommand{\be}{\begin{equation}}
\newcommand{\ee}{\end{equation}}

\begin{document}

\title{Falsifying Field-based Dark Energy Models}

\author{Genly Leon, Yoelsy Leyva, Emmanuel N. Saridakis, \\ Osmel Martin and Rolando Cardenas}

\maketitle

\noindent Department of Mathematics, \\Universidad Central de Las
Villas, Santa Clara \enskip CP 54830, Cuba \\  Department of
Physics,
\\Universidad Central de Las Villas, Santa Clara \enskip CP 54830,
Cuba and \\ Department of Physics, \\ University of Athens,
GR-15771 Athens, Greece \\ Email: genly@uclv.edu.cu,
yoelsy@uclv.edu.cu, msaridak@phys.uoa.gr, osmel@uclv.edu.cu and
rcardenas@uclv.edu.cu

\begin{abstract}
We survey the application of specific tools to distinguish amongst
the wide variety of dark energy models that are nowadays under
investigation. The first class of tools is more mathematical in
character: the application of the theory of dynamical systems to
select the better behaved models, with appropriate attractors in
the past and future. The second class of tools is rather physical:
the use of astrophysical observations to crack the degeneracy of
classes of dark energy models. In this last case the observations
related with structure formation are emphasized both in the linear
and non-linear regimes. We exemplify several studies based on our
research, such as quintom and quinstant dark energy ones. Quintom
dark energy paradigm is a hybrid construction of quintessence and
phantom fields, which does not suffer from fine-tuning problems
associated to phantom field and additionally it preserves the
scaling behavior of quintessence. Quintom dark energy is motivated
on theoretical grounds as an explanation for the crossing of the
phantom divide, i.e. the smooth crossing of the dark energy state
equation parameter below the value -1. On the other hand,
quinstant dark energy is considered to be formed by quintessence
and a negative cosmological constant, the inclusion of this later
component allows for a viable mechanism to halt acceleration. We
comment that the quinstant dark energy scenario gives good
predictions for structure formation in the linear regime, but
fails to do that in the non-linear one, for redshifts larger than
one. We comment that  there might still be some degree of
arbitrariness in the selection of the best dark energy models.
\end{abstract}

\tableofcontents

\section{Introduction}

The current accelerated expansion of our universe has been one of
the most active fields in modern cosmology. Many cosmological
models have been proposed to interpret this mysterious phenomenon,
see e.g. \cite{Copeland2006c, Caldwell:2009ix} for recent reviews.
The simplest candidate is a positive cosmological constant
$\Lambda$ \cite{Carroll:1991mt, Sahni:1999gb}. It is well-known
that its interpretation as the vacuum energy is problematic
because of its exceeding smallness \cite{Padmanabhan:2002ji}.
Notwithstanding its observational merits, the $\Lambda$CDM
scenario is seriously plagued by the well known coincidence and
fine tuning problems \cite{Sahni:2002kh} which are the main
motivations to look for alternative models.

Dark energy (DE) models with two scalar fields ({quintessence} and
phantom) have settled out explicitly and named quintom models
\cite{quintom,quintom1,quintom2,quintom3,quintom4,Guo:2004fq,Zhang:2005eg,Wei:2005fq,Wei:2005nw,Lazkoz:2006pa,stringinspired,stringinspired1,stringinspired2,Cai:2008gk,Saridakis:2009uu,Lazkoz2007,arbitrary,arbitrary1,arbitrary2}.
The quintom paradigm is a hybrid construction of a quintessence
component, usually modelled by a real scalar field that is
minimally coupled to gravity, and a phantom field: a real scalar
field --minimally coupled to gravity-- with negative kinetic
energy. Let us define the equation of state parameter of any
cosmological fluid as $w\equiv\text{pressure}/\text{density}$. The
simplest model of dark energy (vacuum energy or cosmological
constant) is assumed to have $w =-1$. A key feature of
quintom-like behavior is the crossing of the so called phantom
divide, in which the equation of state parameter crosses through
the value $w=-1.$ \footnote{In section \ref{observ} we refer
briefly to observational evidence in favor the quintom DE model.}
Quintom behavior (i.e., the $w=-1$ crossing) has been investigated
in the context of h-essence cosmologies
\cite{Wei:2005fq,Wei:2005nw}; in the context of holographic dark
energy
\cite{holographic,holographic1,holographic2,holographic3,holographic4};
inspired by string theory
\cite{stringinspired,stringinspired1,stringinspired2}; derived
from spinor matter \cite{Cai:2008gk}; for arbitrary potentials
\cite{Lazkoz2007,arbitrary,arbitrary1,arbitrary2}; using
isomorphic models consisting of three coupled oscillators, one of
which carries negative kinetic energy (particularly for
investigating the dynamical behavior of massless
quintom)\cite{setare1}. The crossing of the phantom divide is also
possible in the context of scalar tensor theories
\cite{Elizalde2004,Apostolopoulos:2006si,Bamba:2008xa,Bamba:2008hq,Setare:2008mb}
as well as in modified theories of gravity \cite{Nojiri:2006ri}.

The cosmological evolution of quintom model with exponential
potential has been examined, from the dynamical systems viewpoint,
in \cite{Guo:2004fq} and \cite{Zhang:2005eg,Lazkoz:2006pa}. The
difference between \cite{Guo:2004fq} and
\cite{Zhang:2005eg,Lazkoz:2006pa} is that in the second case the
potential considers the interaction between the conventional
scalar field and the phantom field. In \cite{Zhang:2005eg} it had
been proven that in the absence of interactions, the solution
dominated by the phantom field should be the attractor of the
system and the interaction does not affect its attractor behavior.
In \cite{Lazkoz:2006pa} the case in which the interaction term
dominates against the mixed terms of the potential, was studied.
It was proven there, that the hypothesis in \cite{Zhang:2005eg} is
correct only in the cases in which the existence of the phantom
phase excludes the existence of scaling attractors (in which the
energy density of the quintom field and the energy density of DM
are proportional). Some of this results were extended in
\cite{Lazkoz2007}, for arbitrary potentials. There it was settled
down under what conditions on the potential it is possible to
obtain scaling regimes. It was proved there, that for arbitrary
potentials having asymptotic exponential behavior, scaling regimes
are associated to the limit where the scalar fields diverge.  Also
it has been proven that the existence of phantom attractors in
this framework is not generic and consequently the corresponding
cosmological solutions lack the big rip singularity.

In the first part of the chapter we investigate basic cosmological
observables of quintom paradigm. We perform the cosmological
perturbations analysis of quintom model for independent quadratic
potentials. We investigate the evolution of quintom cosmology with
exponential potentials in a background of a comoving perfect uid.
First, we review the at FRW subcase (with dust background). Then,
we consider both negative and positive curvature FRW models. We
construct two dynamical systems, one adapted to negative curvature
and the other adapted to positive curvature. We characterize the
critical points of the resulting systems. By devising well-defined
monotonic functions we get global results for ever expanding and
contracting models. We find the existence of orbits starting from
and recollapsing to a singularity (given by a massless scalar
field cosmology) for positive curvature models. There is also a
closed FRW solution with no scalar field starting from a big-bang
and recollapsing to a ``big-crunch". We have determined conditions
for the existence of different types of global attractors.
Furthermore, our monotonic functions rule out periodic orbits,
recurrent orbits or homoclinic orbits. We comment about the
interplay between dynamical analysis and observational checking as
tools for discriminate among different quintom proposals.

A large variety of dark energy models suffers from the eternal
acceleration problem, due to the exponential de-Sitter expansion.
One of the consequences of the eternal acceleration is that a
cosmic horizon appears (see e.g. \cite{Cardone2008} for a further
discussion). This problem is not strictly related to $\Lambda$,
should we replace $\Lambda$ with a quintessence scalar field, the
universe should still be eternally accelerated finally reaching a
de Sitter phase and hence again a finite cosmic horizon. In the
second part of the chapter we explore from both observational
testing and dynamical systems perspective a theoretical model to
address the horizon problem. We consider an effective dark energy
fluid as a source of the accelerated expansion. We follow a model
presented by some of us \cite{RolLambda, Cardone2008} whose dark
energy component is the sum of a negative cosmological constant
and a quintessence scalar field evolving under the action of an
exponential potential\footnote{another point of view of the
composite dark energy models can be found in \cite{Grande:2006gb,
Grande:2006qi, Grande:2006nn, Grande:2007dk} where the cosmon
model is introduced }. As a result, although the model is
presently accelerating, eternal acceleration disappears and the
universe ends in a Big Crunch like singularity in a finite time.
Motivated by these theoretical virtues, we further explore this
model from the observational point of view in order to see whether
a negative $\Lambda$ is indeed compatible with the astrophysical
data at hand. We conclude that a negative $\Lambda$ is indeed
allowed and could represent a viable mechanism to halt eternal
acceleration. We also explore the predictions of this class of
model concerning the structure formation in the Universe. We
conclude  that this model give good predictions for structure
formation in the linear regime, but fail to do so in the
non-linear.

\section{Observational Evidence for Quintom Dark Energy Paradigm}\label{observ}

In this section we are going to refer briefly on the observational
evidence that favor the quintom DE model.

\subsection{Basic observables}\label{basicobserv}

In this subsection we examine the basic observational quantity,
which is the dark energy (DE) Equation-of-State (EoS) parameter.
In 2004,  supernovae Ia data were accumulated, opening the road to
constraint imposition on the time variation of DE EoS. In
\cite{Huterer:2004ch}  uncorrelated and nearly model independent
band power estimates (basing on the principal component analysis
\cite{Huterer:2002hy}) of the EoS of DE and its density as a
function of redshift were presented, by fitting to the SNIa data.
Quite unexpectedly, they found marginal ($2\sigma$) evidence for
$w(z)<-1$ at $z < 0.2$, which is consistent with other results in
the literature \cite{Wang:2004py, Alam:2004jy, Wang:2003gz,
Alam:2003fg, Padmanabhan:2002vv, Zhu:2004cu}.

The aforementioned result implied that the EoS of DE could indeed
vary with time. Therefore, one could use a suitable
parametrization of $w_{\rm DE}$ as a function of the redshift $z$,
in order to satisfactory describe such a behavior. There are two
well-studied parametrizations. The first (ansatz A) is:
\begin{equation}
\label{ansatzA}
 w_{\rm DE}=w_0+w'z~,
\end{equation}
where $w_0$ the DE EoS at present and $w'$ an additional
parameter. However, this parametrization is only valid at low
redshift, since it suffers from severe divergences at high ones,
for example ate the last scattering surface $z\sim1100$.
Therefore, a new, divergent-free ansatz (ansatz B) was proposed
\cite{Chevallier:2000qy, Linder:2002et}:
\begin{equation}
\label{ansatzB}
 w_{\rm DE}=w_0+w_1(1-a)=w_0+w_1\frac{z}{1+z}~,
\end{equation}
where $a$ is the scale factor and $w_1=-dw/da$. This
parametrization exhibits a very good behavior at high redshifts.

In \cite{Feng:2004ad} the authors used the ``gold" sample of 157
SNIa, the low limit of cosmic ages and the HST prior, as well as
the uniform weak prior on $\Omega_mh^2$, to constrain the free
parameters of above two DE parameterizations.
\begin{figure}[ht]
\begin{center}
\mbox{\epsfig{figure=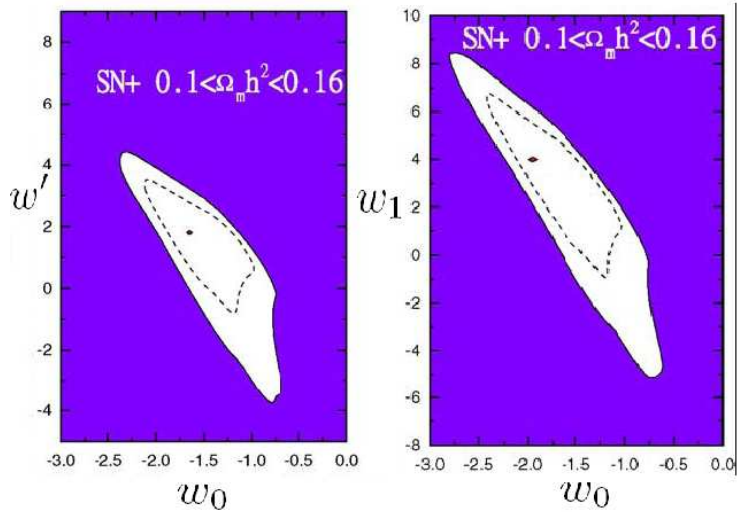,width=9.9cm,angle=0}}
\caption{(Color online) {\it Two-dimensional contour-plots of the
DE equation-of-state parameters, in two different
parameterizations and using SNIa data. The left graph corresponds
to ansatz A (expression (\ref{ansatzA})) and the right graph are
to ansatz B (expression (\ref{ansatzB})). From Ref.
\cite{Feng:2004ad}.}}
 \label{fig4}
\end{center}
\end{figure}
As can be seen in Fig.\ref{fig4} they found that the data seem to
favor an evolving DE with the EoS being below $-1$ around the
present epoch, while it was in the range $w>-1$ in the near
cosmological past. This result holds for both parametrizations
(\ref{ansatzA}),(\ref{ansatzB}), and in particular the best fit
value of the EoS at present is $w_0<-1$, while its ``running''
coefficient is larger than $0$.

\begin{figure}[ht]
\begin{center}
\mbox{\epsfig{figure=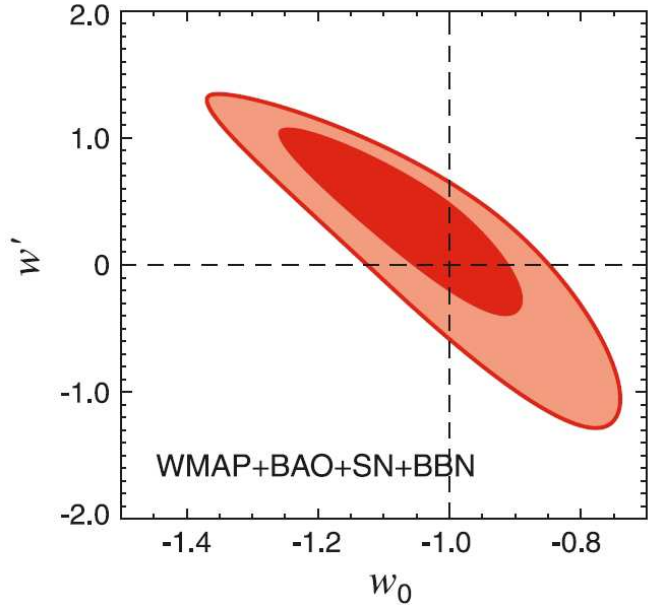,width=8.9cm,angle=0}}
\caption{(Color online) {\it Two-dimensional contour-plot of the
DE equation-of-state parameters, in parameterization ansatz B
(expression (\ref{ansatzB})), and using WMAP, BAO, SNIa data. From
Ref. \cite{Komatsu:2008hk}.}}
 \label{fig5}
\end{center}
\end{figure}

 Apart from the SNIa data, CMB and
LSS data  can be also used to study the variation of EoS of DE. In
\cite{Hannestad:2004cb}, the authors used the first year WMAP,
SDSS and 2dFGRS data to constrain different DE models. They indeed
found that evidently the data favor a strongly time-dependent
$w_{\rm DE}$, and this result is consistent with similar project
of the literature
\cite{Xia:2004rw,Xia:2005ge,Xia:2006cr,Zhao:2006bt,Xia:2006rr,
Xia:2006wd,Zhao:2006qg,Wang:2007mza,Wright:2007vr,Li:2008cj}.
Using the latest 5-year WMAP data, combined with SNIa and BAO
data, the constraints on the DE parameters of ansatz B are:
$w_0=-1.06\pm0.14$ and $w_1=0.36\pm0.62$ \cite{Komatsu:2008hk,
Xia:2008ex, Li:2008vf}, and the corresponding contour plot is
presented in Fig.\ref{fig5}.

In conclusion, as can be observed, the current observational data
mildly favor $w_{\rm DE}$ crossing the phantom divide during the
evolution of universe.

Let us make some comments here. First of all, we mention that the
above results can also fit the basic $\Lambda$CDM paradigm, where
dark energy is attributed to the simple cosmological constant.
Thus, many authors believe that according to data resolution we
can still trust the $\Lambda$CDM paradigm, and thus there is no
need to introduced additional and more complex mechanisms. The
second comment is the following: even if we accept that the
results seem to favor a DE EoS below $-1$ at present, this does
not necessarily means that a two field explanation (one canonical
and one phantom, i.e the basic quintom model) is automatically
justified. One can still result to $w_0<-1$ through many different
frameworks including modified gravity, braneworld constructions,
stringy or strong-inspired models, spinor models etc
\cite{Cai:2009zp}. Thus, in order to distinguish between these
alternatives, one has to find more complicated signatures of the
two-filed quintom model, apart from the simple observable of DE
EoS. One step towards this direction is to investigate the
perturbation spectrum of two-field quintom model, and then examine
its relation to observations.

\subsection{Perturbation theory and current observational
constraints}\label{sec:perturbation}

In this subsection we study the perturbations of two-field quintom
DE paradigm and the effects of these perturbations on the current
observations. Additionally, since it is important to check the
consistency of this model at the classical level, it requires us
to analyze the behavior of perturbations when the EoS crosses the
cosmological constant boundary \cite{Zhao:2005vj}.

\subsubsection{Analysis of perturbations in quintom
cosmology}\label{sec:quintompert}

 In the following
discussion on the quintom perturbations we will restrict ourselves
to the two-field quintom model, with a Lagrangian:
\begin{equation}\label{quintomlag}
    \mathcal{L}=\mathcal{L}_{Q}+\mathcal{L}_{P},
\end{equation}
 where \begin{equation}\label{qlag}
    \mathcal{L}_{Q}=\frac{1}{2}\partial_{\mu}\phi_{1}\partial^{\mu}\phi_{1}-V_{1}(\phi_{1})
\end{equation}
describes the canonical (quintessence) component, and
\begin{equation}
\mathcal{L}_{P}=-\frac{1}{2}\partial_{\mu}\phi_{2}\partial^{\mu}\phi_{2}-V_{2}(\phi_{2})
\end{equation}
the phantom one. The equations of motion for the two scalar
 fields $ \phi_i ( i=1, 2) $ read
\begin{equation}\label{phiEOM}
      \ddot{\phi_i} + 2 \mathcal{H} \dot{\phi_i} \pm a^2 \frac{\partial  V_i}{\partial \phi_i}=
      0,
\end{equation}
where the positive sign is for the quintessence and the minus sign
for the phantom field. Although in general the two scalar fields
could be coupled with each other, here for simplicity we neglect
these interactions.

Now, for a complete study on the perturbations, apart from the
fluctuations of the fields, one has to consider also  the metric
perturbations. In the conformal Newtonian gauge the perturbed
metric writes
\begin{equation}\label{lineelecon}
ds^{2}=a^2(\tau)[(1+2\Psi)d\tau^{2} - (1-2\Phi)dx^{i}dx_{i}].
\end{equation}
Using the notation of \cite{Ma:1995ey}, the perturbation equations
satisfied by each of the quintom components are:
\begin{eqnarray}
    \dot\delta_{i}&=&-(1+w_{i})(\theta_{i}-3\dot{\Phi})
    -3\mathcal{H}\left(\frac{\delta
    P_{i}}{\delta\rho_{i}}-w_{i}\right)\delta_{i}~,\ \ \ \ \label{dotdelta}\\
\dot\theta_{i}&=&-\mathcal{H}(1-3w_{i})\,\theta_{i}-\frac{\dot{w_{i}}}{1+w_{i}}\,\theta_{i}
+k^{2}\left(\frac{\delta
    P_{i}/\delta\rho_{i}}{1+w_{i}}\,\delta_{i}-\sigma_{i} + \Psi\right)~,\label{dottheta}
\end{eqnarray}
 where
\begin{equation}
\theta_{i}=(k^{2}/\dot{\phi}_{i}) \delta\phi_{i},
~~~\sigma_{i}=0~,
\end{equation}
\begin{equation}
w_{i}=\frac{P_{i}}{\rho_{i}}~,
\end{equation}
 and
\begin{equation}
\label{prho2}
    \delta P_{i}=\delta\rho_{i}-2V_{i}'\delta\phi_i=\delta\rho_{i}+ \frac{\rho_{i}\theta_{i}}
      {k^{2}}\left[3\mathcal{H}(1-w_{i}^{2})+\dot{w_{i}}\right]~.
\end{equation}
Thus, combining Eqs. (\ref{dotdelta}), (\ref{dottheta}) and
(\ref{prho2}), we obtain
\begin{eqnarray}
\dot\theta_{i}&=&2\mathcal{H}\theta_{i}+\frac{k^{2}}{1+w_{i}}\delta_{i}+k^2\Psi~,\label{theta2}\\
    \dot\delta_{i}&=&-(1+w_{i})(\theta_{i}-3\dot{\Phi})
  -3\mathcal{H}(1-w_{i})\delta_{i}
       -3\mathcal{H}\left[\frac{\dot w_{i}+3
    \mathcal{H}(1-w_{i}^{2})}{k^{2}}\right]\theta_{i}.\ \ \ \label{delta2}
\end{eqnarray}
Since the simple two-field quintom model is essentially a
combination of a quintessence and a phantom field, one obtains the
perturbation equations by combining the aforementioned equations.
The corresponding variables for the quintom system are
\begin{equation}\label{wq}
    w_{quintom}=\frac{\sum_i P_{i}}{\sum_i \rho_{i}}~,
\end{equation}
\begin{equation}\label{delta}
   \delta_{quintom}=\frac{\sum_i\rho_{i}\delta_{i}}{\sum_i
   \rho_{i}}~,
\end{equation} and
\begin{equation}\label{theta}
    \theta_{quintom}=\frac{\sum_i(\rho_{i}+p_{i})\theta_{i}}{\sum_i
    (\rho_{i}+P_{i})}~.
\end{equation}
Note that for the quintessence component, $-1\leq w_{1}\leq 1$,
while for the phantom component, $w_{2}\leq-1$.

The two-field quintom model is characterized by the potentials
$V_i$. Lets us not consider the simplified case of quadratic
potentials $V_i(\phi_i)=\frac{1}{2} m^2_i \phi^2_i$.
 In general the perturbations of $\phi_i$ arise from
two origins, namely from the adiabatic and the isocurvature modes.
Using instead of $\delta_i$ the gauge invariant variable
$\zeta_i=-\Phi-\mathcal{H}\frac{\delta \rho_i}{\dot{\rho_i}}$, and
in addition the relation $\Phi=\Psi$ in a universe without
anisotropic stress, the equations (\ref{delta2}) and
(\ref{theta2}) can be rewritten as,
\begin{eqnarray}
    \dot\zeta_{i}&=&-\frac{\theta_{i}}{3}-C_i\left(\zeta_i+\Phi+\frac{\mathcal{H}}{k^2}\theta_i\right)
    \label{mdelta2}\\
\dot\theta_{i}&=&2\mathcal{H}\theta_{i}+k^{2}(3\zeta_i+4\Phi)~,
\label{mtheta2}
\end{eqnarray}
where
\begin{eqnarray}
\label{definec}
 C_i=\frac{\dot
w_i}{1+w_i}+3\mathcal{H}(1-w_i)=\partial_0[\ln(a^6|\rho_i+p_i|)].
\end{eqnarray}
In these expressions $\zeta_i$ is the curvature perturbation on
the uniform-density hypersurfaces for the $i$-component of the
universe \cite{Wands:2000dp}. Usually, the isocurvature
perturbations of $\phi_i$ are characterized by the differences
between the curvature perturbation of the uniform-$\phi_i$-density
hypersurfaces and that of the uniform-radiation-density
hypersurfaces,
\begin{eqnarray}
 S_{ir}\equiv 3(\zeta_i-\zeta_r),
\end{eqnarray}
where the subscript $r$ stands for radiation. Here we assume there
are no matter isocurvature perturbations, and thus
$\zeta_m=\zeta_r$. Eliminating $\zeta_i$ in equations
(\ref{mdelta2}) and (\ref{mtheta2}), we obtain a second order
equation for $\theta_i$, namely
\begin{equation}
\label{2theta}
  \ddot
\theta_i+(C_i-2\mathcal{H})\dot\theta_i+(C_i{\cal H}-2\dot{\cal
H}+k^2)\theta_i=k^2(4\dot\Phi+C_i\Phi).
\end{equation}
The general solutions of this inhomogeneous differential equation,
is the sum of the general solution of its homogeneous part with a
special integration.  In the following, we will show that the
special integration corresponds to the adiabatic perturbation.

As it assumed, before the era of DE domination, the universe was
dominated by either radiation or dark matter. The perturbation
equations for these background fluids read:
\begin{eqnarray}
& &\dot\zeta_f=-\theta_f/3~,\nonumber\\
&
&\dot\theta_f=-\mathcal{H}(1-3w_f)\theta_f+k^2[3w_f\zeta_f+(1+3w_f)\Phi]~.\
\ \ \ \ \ \ \
\end{eqnarray}
From the Poisson equation
\begin{eqnarray}
-\frac{k^2}{\mathcal{H}^2}\Phi=\frac{9}{2}\sum_{\alpha}\Omega_{\alpha}(1+w_{\alpha})
\left(\zeta_{\alpha}+\Phi+\frac{\mathcal{H}}{k^2}\theta_{\alpha}\right)\simeq\frac{9}{2}(1+w_{f})
\left(\zeta_{f}+\Phi+\frac{\mathcal{H}}{k^2}\theta_{f}\right)~,
\end{eqnarray}
on large scales we approximately acquire:
\begin{eqnarray}
\Phi\simeq-\zeta_f-\frac{\mathcal{H}}{k^2}\theta_{f}.
\end{eqnarray}
Therefore, combining the equations above with ${\cal
H}=2/[(1+3w_f)\tau]$, we get (note that numerically $\theta_f\sim
\mathcal{O}(k^2)\zeta_f$)
\begin{eqnarray}
& &\zeta_f=-\frac{5+3w_f}{3(1+w_f)}\Phi={\rm const.}~,\nonumber\\
& &\theta_f=\frac{k^2 (1+3w_f)}{3(1+w_f)}\Phi\tau~.
\end{eqnarray}
Therefore,  from (\ref{2theta}) we observe that there is a special
solution  which on large scales  it is given approximately by
\begin{eqnarray}
\theta_i^{ad}=\theta_f,
 \end{eqnarray}
  while
(\ref{mtheta2}) leads to
  \begin{eqnarray}
\zeta_i^{ad}=\zeta_f.
\end{eqnarray}
 This indicates  that the special
integration of (\ref{2theta}) corresponds to the adiabatic
perturbation. Hence, concerning the isocurvature perturbations of
$\phi_i$, we can consider only the solution to the homogeneous
part of (\ref{2theta}),
  \begin{eqnarray}
  \label{22theta}
 \ddot
\theta_i+(C_i-2\mathcal{H})\dot\theta_i+(C_i\mathcal{H}-2\dot{\cal
H}+k^2)\theta_i=0~.
\end{eqnarray}
These solutions are represented by $\theta_i^{iso}$ and
$\zeta_i^{iso}$. The relation between them is
  \begin{eqnarray}
  \label{iso}
\zeta_i^{iso}=\frac{\dot\theta_{i}^{iso}-2\mathcal{H}\theta_{i}^{iso}}{3k^{2}}~.
\end{eqnarray}
 Since the general solution of $\zeta_i$ is \be
\zeta_i=\zeta_i^{ad}+\zeta_i^{iso}=\zeta_r+\zeta_i^{iso}~, \ee the
isocurvature perturbations are simply $S_{ir}=3\zeta_i^{iso}$.

In order to solve (\ref{22theta}), we need to know the forms of
$C_i$ and $\mathcal{H}$ as functions of time $\tau$. For this
purpose, we solve the background equations (\ref{phiEOM}). During
the radiation dominated period, $a=A\tau~,~\mathcal{H}=1/\tau$ and
we thus we have
 \be
 \label{phi1}
\phi_1=\tau^{-1/2}\left[A_{1}J_{1/4}\left(\frac{A}{2}m_1\tau^2\right)+A_{2}J_{-1/4}\left(\frac{A}{2}m_1\tau^2\right)\right]~,
\ee and \be\label{phi2}
\phi_2=\tau^{-1/2}\left[\tilde{A}_{1}I_{1/4}\left(\frac{A}{2}m_2\tau^2\right)+\tilde{A}_{2}I_{-1/4}\left(\frac{A}{2}m_2\tau^2\right)\right]~,
\ee
 respectively, where $A$, $A_i$ and $\tilde{A}_i$ are
constants, $J_{\nu}(x)$ is the $\nu$th order  Bessel function and
$I_{\nu}(x)$ is the $\nu$th order modified Bessel function. Since
the masses are usually small in comparison with the expansion rate
of the early universe $m_i\ll \mathcal{H}/a$, we can approximate
the (modified) Bessel functions as $J_{\nu}(x)\sim
x^{\nu}(c_1+c_2x^2)$ and $I_{\nu}(x)\sim
x^{\nu}(\tilde{c}_1+\tilde{c}_2x^2)$.
 We mention that $J_{-1/4}$ and
$I_{-1/4}$ are divergent when $x\rightarrow 0$. Given these
arguments we can see that it requires large initial values of
$\phi_i$ and $\dot\phi_i$ if $A_2$ and $\tilde{A}_2$ are not
vanished. Imposing small initial values, which is the natural
choice if the DE fields are assumed to survive after inflation,
only $A_1$ and $\tilde{A}_1$ modes exist, so $\dot\phi_i$ will be
proportional to $\tau^3$ in the leading order. Thus, the
parameters $C_i$ in (\ref{definec}) will be $C_i=10/\tau$ (we have
used $|\rho_i+p_i|=\dot\phi_i^2/a^2$). So, we acquire the solution
of (\ref{22theta}), \be
\theta_i^{iso}=\tau^{-4}[D_{i1}\cos(k\tau)+D_{i2}\sin(k\tau)].
 \ee
 Therefore, $\theta_i^{iso}$ presents an oscillatory behavior,
  with an amplitude damping with the
expansion of the universe. This fact leads the isocurvature
perturbations $\zeta_i^{iso}$ to decrease rapidly. If we choose
large initial values for $\phi_i$ and $\dot \phi_i$, $A_2$ and
$\tilde{A}_2$ modes are present, $\dot\phi_i$ will be proportional
to $\tau^{-2}$ in the leading order and $C_i=0$. Now the solution
of (\ref{22theta}) is:
 \be
\theta_i^{iso}=\tau[D_{i1}\cos(k\tau)+D_{i2}\sin(k\tau)]~.
 \ee
That is, $\theta_i^{iso}$ will oscillate with an increasing
amplitude, so $\zeta_i^{iso}$ remains constant on large scales.

Similarly, during matter dominated era,
$a=B\tau^2~,~\mathcal{H}=2/\tau$, and thus the solutions for the
fields $\phi_i$ respectively read
 \be\label{phi11}
\phi_1=\tau^{-3}\left[B_{1}\sin\left(\frac{B}{3}m_1\tau^3\right)+B_{2}\cos\left(\frac{B}{3}m_1\tau^3\right)\right]
\ee and \be\label{phi22}
\phi_2=\tau^{-3}\left[\tilde{B}_{1}\sinh\left(\frac{B}{3}m_2\tau^3\right)+\tilde{B}_{2}\cosh\left(\frac{B}{3}m_2\tau^3\right)\right].
\ee
 Therefore, we do get the same conclusions with the analysis
for the radiation dominated era. Firstly, choosing small initial
values at the beginning of  matter domination, we deduce that the
isocurvature perturbations in $\phi_i$ will decrease with time. On
the contrary, for large initial values, the isocurvature
perturbations remain constant on large scales. This behavior was
expected, since in the case of large initial velocity the energy
density of the scalar field is dominated by the kinetic term and
it behaves like a fluid with $w=1$, and thus its isocurvature
perturbation remains constant on large scales. However, on the
other hand, the energy density of the scalar field will be
dominated by the potential energy due to the slow rolling, that is
it will behave like a cosmological constant and thus there are
only tiny isocurvature perturbations in it.

In summary, we have seen that the isocurvature perturbations in
quintessence-like or phantom-like field under quadratical
potentials decrease or remain constant at large scales, depending
on the initial velocities. In other words, the isocurvature
perturbations are stable on large scales, with their amplitude
being proportional to the value of Hubble parameter evaluated
during the period of inflation $H_{inf}$ (if indeed their quantum
nature originates from inflation). In the case of a large
$H_{inf}$, the isocurvature dark energy perturbations can be
non-negligible and thus they will contribute to the observed CMB
anisotropy \cite{Kawasaki:2001bq,Moroi:2003pq}. However, in the
cases analyzed in this subsection, these isocurvature
perturbations are negligible. Firstly, as mentioned above, large
initial velocities are not possible if we desire the quintom
fields to survive after inflation. Furthermore, even if the
initial velocities are large at the beginning of the radiation
domination, they will be reduced to a small value due to the small
masses and the damping effect of Hubble expansion.

 In conclusion, we deduce
that the contributions of DE isocurvature perturbations are not
very large \cite{Gordon:2005ti} and thus for simplicity we assume
that $H_{inf}$ is small enough in order to make the isocurvature
contributions negligible. Therefore, it is safe to focus only in
the effects of the adiabatic perturbations of the quintom model.

\subsubsection{Signatures of perturbations in quintom
scenario}\label{sec:quintomsignal}

Let us now investigate the observational signatures of
perturbations in quintom scenario. For this shake we use the
perturbation equations(\ref{delta}) and (\ref{theta}), and we are
based on the code of CAMB \cite{Lewis:1999bs}. For simplicity we
impose a flat geometry as a background, although this is not
necessary. Moreover, we assume the fiducial background parameters
to be $\Omega_{b}=0.042, \Omega_{DM}=0.231, \Omega_{DE}=0.727$,
where $b$ stands for baryons, $DM$ for dark matter and $DE$ for
dark energy, while today's Hubble constant is fixed at
$H_{0}=69.255$ km/s Mpc$^{-2}$. We will calculate the effects of
perturbed quintom on CMB and LSS.

In the two-field quintom model  there are two parameters, namely
the quintessence and phantom masses. When the quintessence mass is
larger than the Hubble parameter, the field starts to oscillate
and consequently one obtains an oscillating quintom. In the
numerical analysis we will fix the phantom mass to be $m_P\sim 2.0
\times 10^{-60} M_{pl}$, and we vary the quintessence mass with
the typical values being $m_Q=10^{-60} M_{pl}$ and $4 \times
10^{-60} M_{pl}$ respectively.\\

\underline{Oscillatory Quintom}\\

\begin{figure}[!]
\mbox{\epsfig{figure=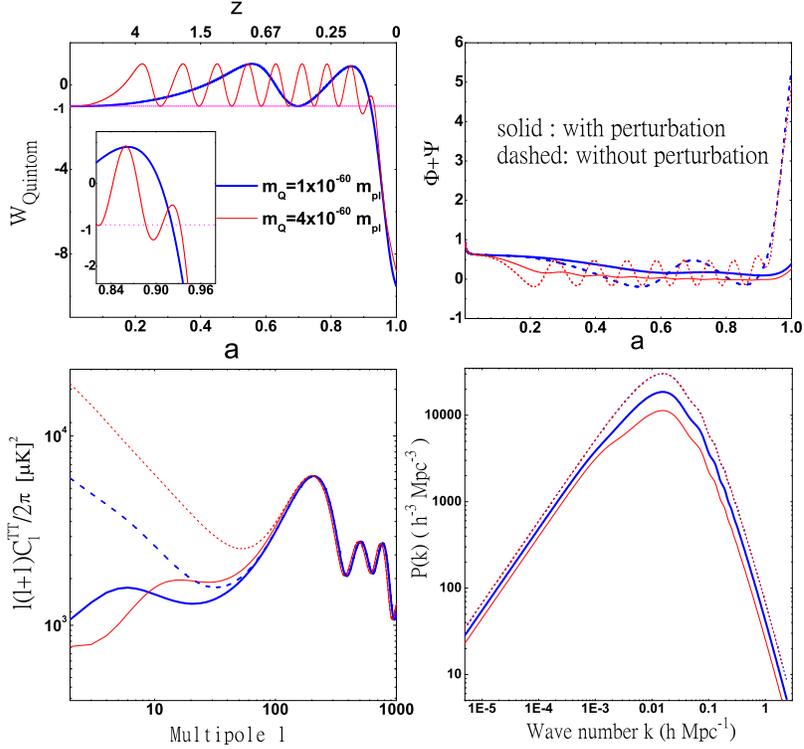,width=12.6cm,angle=0}}
\caption{(Color online) {\it Effects of the two-field oscillating
quintom on the observables. The phantom mass is fixed at
$2.0\times10^{-60} M_{pl}$ and the  quintessence mass at $10^{-60}
M_{pl}$ (thicker line) and $4.0\times10^{-60} M_{pl}$ (thinner
line) respectively. The upper right graph depicts the evolution of
the metric perturbations $\Phi+\Psi$ of the two models,  with
(solid lines) and without(dashed lines) DE perturbations. The
scale is $k\sim10^{-3}$ Mpc$^{-1}$. The lower left graph shows the
CMB effects and the lower right one delineates the effects on the
matter power-spectrum, with (solid lines) and without (dashed
lines) DE perturbations. From Ref. \cite{Zhao:2005vj}. }}
 \label{ZhaoFig1}
\end{figure}

In Fig. \ref{ZhaoFig1} we depict the equation-of-state parameters
as a function of the scale factor, for the aforementioned two
parameter-sets, and additionally their corresponding effects on
observations. We clearly observe the quintom oscillating behavior
as the mass of quintessence component increases. After reaching
the $w=-1$ pivot for several times, $w$ crosses $-1$ consequently
with the phantom-component domination in dark energy. As a result,
the quintom fields modifies the metric perturbations: $\delta
g_{00}=2a^{2}\Psi,\delta g_{ii}=2a^{2}\Phi \delta_{ij}$ and
consequently they contribute to the late-time Integrated
Sachs-Wolfe (ISW) effect. The ISW effect is an integrant of
$\dot{\Phi}+\dot{\Psi}$ over conformal time and wavenumber $k$.
The above two specific quintom models yield quite different
evolving $\Phi+\Psi$ as shown in the right panel of Fig.
\ref{ZhaoFig1}, where the scale is $k\sim 10^{-3}$ Mpc$^{-1}$. As
we  can see, the late time ISW effects differ significantly when
DE perturbations are taken into account(solid lines) or not(dashed
lines).

ISW effects constitute an important part on large angular scales
of CMB and on the matter power spectrum of LSS. For a constant EoS
of phantom it has been shown that the low multipoles of CMB will
get significantly enhanced when DE perturbations are neglected
\cite{Weller:2003hw}. On the other hand for a matter-like scalar
field, where the EoS is around zero, perturbations will also play
an important role on the large scales of CMB
\cite{Caldwell:1997ii}. Our results on CMB and LSS reflect the two
combined effects of phantom and oscillating quintessence. We
mention that while in the early studies of quintessence effects on
CMB, one could consider a constant $w_{eff}$ instead:
\begin{equation}\label{weff}
 w_{eff}\equiv\frac{\int da \Omega(a) w(a)}{\int da \Omega(a)}~~,
\end{equation}
 this is not enough for the study of effects on SNIa, nor
for CMB, when the EoS of DE has a very large variation with
redshift, such as the model of oscillating quintom considered
above.

\begin{figure}[!]
\begin{center}
\mbox{\epsfig{figure=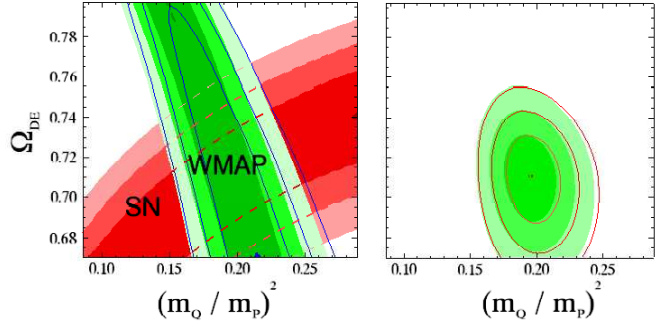,width=9.cm,angle=0}} \caption{(Color
online) {\it 3$\sigma$ WMAP and SNIa constraints on two-field
quintom model, shown together with the best fit values. $m_Q$ and
$m_P$ denote the quintessence and phantom  mass respectively. We
have fixed $m_P\sim 6.2 \times 10^{-61} M_{pl}$ and we have varied
the value of $m_Q$. Left graph: separate WMAP and SNIa
constraints. The green (shaded) area marks the WMAP constraints on
models where DE perturbations have been included, while the blue
area (contour with solid lines) corresponds to the case where DE
perturbations have not been taken into account. Right graph:
combined WMAP and SNIa constraints on the two-field quintom model
with perturbations (green/shaded region) and without perturbations
(red region/contour with solid lines). From Ref.
\cite{Zhao:2005vj}. }}
 \label{ZhaoFig2}
\end{center}
\end{figure}

To analyze the oscillating quintom-model  under the current
observations, we perform a preliminary fitting to the first year
WMAP TT and the TE temperature--polarization cross-power spectrum
as well as the recently released 157 ``Gold" SNIa data
\cite{Riess:2004nr}. Following  \cite{Contaldi:2003zv,
Feng:2003zua} in all the fittings below we fix $\tau=0.17$,
$\Omega_m h^2=0.135$ and $\Omega_b h^2=0.022$,  setting the
spectral index as $n_S=0.95$, and using the amplitude of the
primordial spectrum  as a continuous parameter. In the fittings of
oscillating quintom we've fixed the phantom-mass to be $m_P\sim
6.2 \times 10^{-61} M_{pl}$. Fig. \ref{ZhaoFig2} delineates
3$\sigma$ WMAP and SNIa constraints on the two-field quintom
model, and in addition it shows the corresponding best fit values.
The parameters $m_Q$ and $m_P$ stand for the masses of
quintessence and phantom respectively. In the left graph of
Fig.\ref{ZhaoFig2} we present the separate WMAP and SNIa
constraints. The green(shaded) area is WMAP constraints on models
where DE perturbations have been included, while the blue area
(contour with solid lines) is the corresponding area without DE
perturbations. The perturbations of DE have no effects on the
geometric constraint of SNIa. The right graph shows the combined
WMAP and SNIa constraints on the two-field quintom model with
perturbations (green/shaded region) and without perturbations (red
region/contour with solid lines). We conclude that the confidence
regions indeed present a large difference, if the DE perturbations
 have been taken into account or not.
\\

\underline{Non-oscillatory Quintom}\\

\begin{figure}[!]
\begin{center}
\mbox{\epsfig{figure=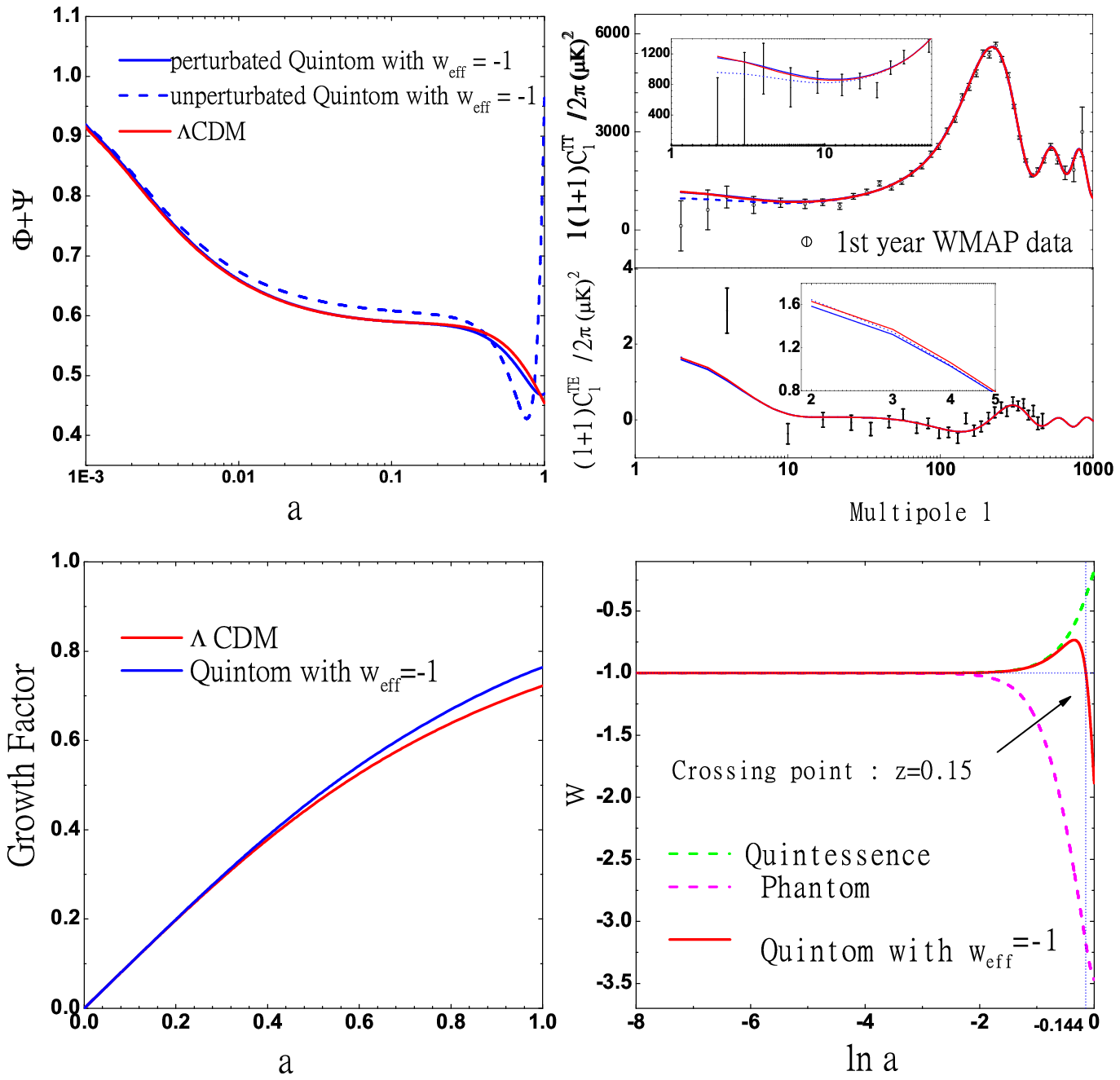,width=12.6cm,angle=0}}
\caption{(Color online) {\it Comparison of the effects of the
two-field quintom model with $w_{eff}=-1$ and of the simple
cosmological constant, in CMB (WMAP), the metric perturbations
$\Phi+\Psi$ (the scale is $k\sim10^{-3}$ Mpc$^{-1}$) and the
linear growth factor. The binned error bars in the upper right
graph are WMAP TT and TE data \cite{Kogut:2003et, Hinshaw:2003ex}.
From Ref. \cite{Zhao:2005vj}. }}
 \label{ZhaoFig3}
\end{center}
\end{figure}

As we have mentioned, the basic observables could also described
by the simple cosmological constant. Thus, in order to distinguish
the quintom model from the cosmological constant, we now consider
a quintom scenario where $w$ crosses $-1$ smoothly without
oscillations. It is interesting to study the effects of this type
of  quintom model, with its effective EoS defined in (\ref{weff})
exactly equal to $-1$, on CMB and matter power spectrum. Indeed,
we have realized such a quintom model in the lower right panel of
Fig. \ref{ZhaoFig3}, which can be easily given in the two-field
model with a lighter quintessence mass. In this example we have
set $m_Q\sim 2.6 \times 10^{-61} M_{pl}, m_P\sim 6.2 \times
10^{-61} M_{pl}$. Additionally, we assume that there is no initial
kinetic energy. The initial value of the quintessence component is
set to $\phi_{1i}=0.226 M_{pl}$, while for the phantom part we
impose $\phi_{2i}= 6.64 \times 10^{-3} M_{pl}$. We find that the
EOS of quintom crosses $-1$ at $z\sim 0.15$, which is consistent
with the latest SNIa results.

The model of quintom, which is mainly favored by current SNIa
only, needs to be confronted with other observations in the
framework of concordance cosmology. Since SNIa offer the only
direct detection of DE, this model is the most promising to be
distinguished from the cosmological constant and other dynamical
DE models which do not get across $-1$, by future SNIa projects on
the low redshift (for illustrations see e.g.
\cite{Huterer:2004ch}). This is also the case for the quintom
model in the full parameter space: it can be most directly tested
in low redshift Type Ia supernova surveys.

In the upper left panel of Fig. \ref{ZhaoFig3} we delineate the
different ISW effects among the cosmological constant (red/light
solid), the quintom model which gives $w_{eff}=-1$ with (blue/dark
solid) and without(blue dashed) perturbations. Similarly to the
previous oscillating case, the difference is very large when
switching off quintom perturbations and much smaller when
including the perturbations. In the upper right panel we find that
the quintom model cannot be distinguished from a cosmological
constant in light of WMAP. The two models  give almost exactly the
same results in CMB TT and TE power spectra when including the
perturbations. We deduce that the difference in CMB is hardly
distinguishable even by cosmic variance.

\subsubsection{Breaking the degeneracy between quintom and cosmological
constant scenarios}

So far we have see that CMB observations cannot distinguish
between a quintom model with $w_{eff} = -1$ and a cosmological
constant. Thus, in order to acquire distinctive signatures, we
have to rely in other observations. To achieve that we need to
consider the physical observables which can be affected by the
evolving $w$ sensitively. In comparison with the cosmological
constant, such a quintom model exhibits a different evolution of
the universe's expansion history, and in particular it gives rise
to a different epoch of matter-radiation equality. The Hubble
expansion parameter becomes:
\begin{equation}\label{H}
    H \equiv \frac{\dot{a}}{a^2}=H_{0}[\Omega_{m}a^{-3}+\Omega_{r}a^{-4}+X]^{1/2}
\end{equation}
where $X$, the energy density ratio of DE between the early times
and today, is quite different between the $quintom$-CDM and
$\Lambda$CDM. In the $\Lambda$CDM scenario $X$ is simply a
constant, while in general for DE models with varying energy
density or EoS we obtain
\begin{equation}\label{omegaquitom}
    X=\Omega_{DE} a^{-3}e^{-3\int w(a)d \ln a}.
\end{equation}
Therefore, the two models will give different Hubble expansion
rates. This is also the case between the quintom model with
$w_{eff} = -1$ in the left panel of Fig. \ref{ZhaoFig3} and a
cosmological constant.

Finally, we mention that different $H$ leads directly to different
behaviors of the growth factor. In particular, according to the
linear perturbation theory all Fourier modes of the matter density
perturbations grow at the same rate, that is the matter density
perturbations are independent of $k$:
\begin{equation}\label{detamk}
    \ddot{\delta}_k + \mathcal{H} \dot{\delta}_k - 4 \pi G a^2 \rho_{\rm
M} \delta_k=0.
\end{equation}
The growth factor $D_{1}(a)$ characterizes the growth of the
matter density perturbations, namely $D_{1}(a)= \delta_k
(a)/\delta_k (a=1)$, and it is normalized to unity today. In the
matter-dominated epoch we have $D_1 (a)=a$. Analytically
$D_{1}(a)$ is often approximated by the Meszaros equation
\cite{Dodelson:2003ft}:
\begin{equation}\label{d1}
    D_{1}(a)=\frac{5\Omega_{m}H(a)}{2H_{0}}\int^{a}_{0}\frac{da'}{(a'H(a')/H_{0})^{3}}.
\end{equation}
Therefore, we can easily observe the difference between the
quintom and cosmological constant scenarios, due to the different
Hubble expansion rates. In particular, one needs to solve
(\ref{detamk}) numerically. In the lower left graph  of Fig.
\ref{ZhaoFig3} we show the difference of $D_1 (a)$ between the
quintom model with $w_{eff} = -1$ and the cosmological constant
one. The difference in the linear growth function is considerably
large in the late time evolution and possibly distinguishable in
future LSS surveys and in weak gravitational lensing (WGL)
observations. WGL has emerged with a direct mapping of cosmic
structures and it has been recently shown that the method of
cosmic magnification tomography can be extremely efficient
\cite{Jain:2003tba, Zhang:2005eb, Zhang:2005pu}, which leaves a
promising future for breaking the degeneracy between quintom and a
cosmological constant.

\section{Exponential quintom: Phase space
analysis}\label{sectionII}

In the following discussion on the quintom phase space analysis we
restrict ourselves to the two-field quintom model, with a
Lagrangian:
\begin{equation}\label{expquintomlag}
    \mathcal{L}=\frac{1}{2}\partial_{\mu}\phi\partial^{\mu}\phi
    -\frac{1}{2}\partial_{\mu}\varphi\partial^{\mu}\varphi-V(\phi,\varphi),
\end{equation}
and we include, also, ordinary matter (a comoving perfect fluid)
in the gravitational action. As in \cite{Lazkoz:2006pa} we
consider here the efective two-field potential \be V=V_0
e^{-\sqrt{6}(m\phi+n\varphi)},\label{ExponentialPot}\ee  where the
scalar field $\phi$ represents quintessence and $\varphi$
represents a phantom field. For simplicity, we assume $m>0$ and
$n>0.$

Quintom (non-conventional) cosmologies with exponential potentials
has been investigated, from the dynamical systems approach, for
instance, in  references
\cite{Guo:2004fq,Zhang:2005eg,Lazkoz:2006pa} (see section
\ref{sectionIII} for a brief review).

We shall consider the Friedmann-Lemaitre-Robertson-Walker (FLRW)
line element: \be ds^2 = - dt^2 + a^2(t)\left(\frac{dx^2}{1-k
x^2}+x^2\left(d{\vartheta}^2 + \sin^2 {\vartheta}\,
d{\varphi}^2\right)\right),\label{fx_FLRW} \ee where  $k =
1,0,-1,$ identifies the three types of FRW universes: closed,
flat, and open, respectively.

The field equations derived from (\ref{fx_FLRW}), are
\begin{eqnarray}
& H^2 - \sfrac16\left(\dot\phi^2-\dot\varphi^2\right)-\sfrac13 V_{\text{eff}}-\sfrac13\rho_{\rm M}=-\sfrac{k}{a^2},\label{Friedmann1}\\
&\dot H=-H^2-\sfrac13\left(\dot\phi^2-\dot\varphi^2\right)+\sfrac13 V_{\text{eff}}-\sfrac16 \rho_{\rm M},\label{Raych}\\
&\dot\rho_{\rm M}=-3 H\rho_{\rm M},\label{consm}\\
&\ddot\phi+3H\dot\phi-\sqrt{6}m V=0,\label{consphi}\\
&\ddot\varphi+3H\dot\varphi+\sqrt{6}n V=0,\label{consvarphi}
\end{eqnarray} where $H =\frac{\dot a(t)}{a(t)}$ denotes de Hubble expansion scalar.

The dot denotes derivative with respect the time $t.$ We consider
a pressureless perfect fluid (dust) as the background matter.

\subsection{Flat FRW subcase}\label{sectionIII}

To investigate the flat models we introduced the same normalized
variables as in \cite{Lazkoz:2006pa}: $(x_\phi,\,x_\varphi,\,y)$,
defined by
\begin{eqnarray} x_\phi=\frac{\dot\phi}{\sqrt{6} H},\;
x_\varphi=\frac{\dot\varphi}{\sqrt{6} H},  y=\frac{\sqrt
V}{\sqrt{3}H}.\label{vvars}
\end{eqnarray}
They are related through the Friedman equation \ref{Friedmann1} by
$x_\phi^2-x_\varphi^2+y^2=1-\frac{\rho_{\rm M}}{3 H^2}\leq 1.$

The dynamics in the space is given by the ordinary differential
equations \cite{Lazkoz:2006pa}:

\bea
&&x_\phi'=\frac{1}{3} \left(3 m y^2+(q-2) x_\phi\right)\label{eqxphi}\\
&&x_\vphi'=-\frac{1}{3} \left(3 n y^2-(q-2) x_\varphi \right)\label{eqxvphi}\\
&&y'=\frac{1}{3} (1+q-3(m x_\phi+n x_\varphi)) y\label{eqy} \eea
defined in the phase space given by \be\Psi=\{{\bf
x}=(x_\phi,x_\vphi,y):0\le
x_\phi^2-x_\vphi^2+y^2\le1\}.\label{phase_space}\ee

Here the prime denotes differentiation with respect to a new time
variable $\tau=\log a^3,$ where $a$ is the scale factor of the
space-time. The deceleration factor $q\equiv-\ddot a a/\dot a^2$
can then be written \be q=\frac{1}{2} \left(3
\left(x_\phi^2-x_\varphi^2-y^2\right)+1\right).\ee

The system (\ref{eqxphi}-\ref{eqy}) admits the critical points
points $O, C_{\pm}, T, P$ reported in \cite{Lazkoz:2006pa}). In
the table \ref{table1} the location, existence and deceleration
factor of the critical points for $m>0$,  $n>0$ and $y>0.$ We use
the notation $\delta=m^2-n^2.$

\begin{table*}[t]
\caption[crit]{Location, existence and deceleration factor of the
critical points for $m>0$,  $n>0$ and $y>0.$ We use the notation
$\delta=m^2-n^2$ (from reference \cite{Lazkoz:2006pa}).}
\label{table1}
\begin{center}
\begin{tabular}{@{\hspace{4pt}}c@{\hspace{14pt}}c@{\hspace{14pt}}c@{\hspace{14pt}}c@{\hspace{18pt}}c@{\hspace{18pt}}c@{\hspace{2pt}}}
\hline
\hline\\[-0.3cm]
Name &$x_\phi$&$x_\vphi$&$y$&Existence&$q$\\[0.1cm]
\hline\\[-0.2cm]
$O$& $0$& $0$& $0$& All $m$ and $n$ &$\ds\frac{1}{2}$\\[0.2cm]
$C_{\pm}$ & $\pm\sqrt{1+{x_{\varphi}^*}^2}$& $x_{\varphi}^*$& $0$&
All $m$ and $n$  &$2$ \\[0.2cm]
${P}$&
$m$& $-n$ &$\sqrt{1-\delta}$&$\delta< 1$&$-1+3\delta$\\[0.2cm]
${T}$& $\ds\frac{\ds m}{\ds{2}{\delta}}$& $-\ds\frac{\ds
n}{\ds{2}{\delta}}$&
$\ds\frac{1}{2\sqrt{\delta}}$&$\delta\ge1/2$&$\ds\frac{1}{2}$\\[0.4cm]
\hline \hline
\end{tabular}
\end{center}
\end{table*}

By analyzing the sign of the real part of the normally-hyperbolic
curves $C_{\pm}$ we get the following results (we are assuming
$m>0$ and $n>0$): \begin{itemize} \item If $m<n,$ $C_+$ contains
an infinite arc parameterized by ${x_\varphi^*}$ such that
${x_\varphi^*}<\frac{-n-m\sqrt{1-\delta}}{\delta}$ that is a local
source. $C_-$ contains an infinite arc parameterized by
${x_\varphi^*}$ such that
${x_\varphi^*}<\frac{-n+m\sqrt{1-\delta}}{\delta}$ that is a local
source. \item If $m=n,$ $C_+$ contains an infinite arc
parameterized by ${x_\varphi^*}$ such that
${x_\varphi^*}<\frac{1-m^2}{2n}$ that is a local source. All of
$C_-$ is a local source. \item If $m>n$ there are two
possibilities
\begin{itemize}
\item If $\delta<1,$ all of $C_-$ is a local source. A finite arc
of $C_+$ parameterized by ${x_\varphi^*}$ such that
$\frac{-n-m\sqrt{1-\delta}}{\delta}<{x_\varphi^*}<\frac{-n+m\sqrt{1-\delta}}{\delta}$
is a local source. \item If $\delta\geq 1,$ no part of $C_+$ is a
local source and all of $C_-$ is a local source.
\end{itemize}
\end{itemize}

Perhaps, the most appealing result in \cite{Lazkoz:2006pa} is
that, by introducing properly defined monotonic functions and by
making some numerical integrations, it was possible to identify
heteroclinic sequences

\begin{itemize}
    \item Case i) For $m<\sqrt{n^2+1/2},$ the point $P$ is a stable node, whereas the point $T$ does not exist. The heteroclinic sequence in this case is $C_{\pm}\longrightarrow O\longrightarrow P.$
    \item Case ii) For $\sqrt{n^2+1/2}<m\leq \sqrt{n^2+4/7},$ the point $T$ is a stable node and the point $P$ is
a saddle. For these conditions the heteroclinic sequence is
$C_{\pm}\longrightarrow O\longrightarrow T \longrightarrow P.$
    \item Case iii) For $\sqrt{n^2+4/7}<m<\sqrt{1+n^2},$ the point T is a spiral node and the point P is a saddle. For these conditions the heteroclinic sequence is the same as in the former case.
    \item Case iv) For $m>\sqrt{1+n^2}$ the point T is a spiral node whereas the point $P$ does not exist. The
heteroclinic sequence in this case is  $C_{-}\longrightarrow
O\longrightarrow T.$
\end{itemize}

From the possibilities listed above, there still the possibility
of other attractors different from phantom ones in the exponential
quintom scenario, particularly scaling attractors ($T$). This fact
is a counterexample of one of the result in \cite{Zhang:2005eg}.
However, we are aware on the small probability that $T$ represents
the actual stage on the universe evolution due this solution is
matter-dominated. Another compelling result in quintom cosmology
is that the existence of monotonic functions in the state space
can rule out periodic orbits, homoclinic orbits, and other complex
behaviour in invariant sets. If so, the dynamics is dominated by
critical points (and possibly, heteroclinic orbits joining it).
Additionally, some global results can be obtained. A similar
approach (i.e., that of devising monotonic functions) for multiple
scalar field cosmologies with matter was used in
\cite{Coley:2003mj,vdhoogen,Coley:1999mj}. However, in that work
they do not consider phantom-like scalar fields, as we do here.

\subsubsection{Analysis at infinity.}

\begin{table*}[t]
\caption[crit]{Location and existence conditions for the critical
points at infinity.} \label{critatinfinity}
\begin{center}
\begin{tabular}{@{\hspace{4pt}}c@{\hspace{14pt}}c@{\hspace{18pt}}c@{\hspace{18pt}}c@{\hspace{2pt}}}
\hline
\hline\\[-0.3cm]
Name &$\theta_1$&$\theta_2$&Existence \\[0.1cm]
\hline\\[-0.2cm]
$P_1^\pm$& $0$& $\pm\frac{\pi}{2}$& always\\[0.2cm]
$P_2^\pm$ & $\pi$ & $\pm\frac{\pi}{2}$&  always\\[0.2cm]
$P_3^\pm$& $\frac{\pi}{4}$& $\pm \cos
^{-1}\left(-\frac{m}{n}\right)$ &$-\pi<\pm\cos
^{-1}\left(-\frac{m}{n}\right)\leq \pi, n\neq 0$\\[0.2cm]
$P_4^\pm$& $\frac{3\pi}{4}$& $\pm \cos
^{-1}\left(\frac{m}{n}\right)$ &$-\pi<\pm \cos
^{-1}\left(\frac{m}{n}\right)\leq \pi, n\neq 0$\\[0.2cm]
$P_5$& $\theta_1^\star$ & $0$
&$0\leq \theta_1^\star\leq \pi$\\[0.2cm]
$P_6$& $\theta_1^\star$ & $\pi$ &$0\leq \theta_1^\star\leq \pi$
\\[0.4cm]
\hline \hline
\end{tabular}
\end{center}
\end{table*}

The numerical experiments in \cite{Lazkoz:2006pa} suggest that
there is an open set of orbits that tends to infinity. Let us
investigate the dynamics at infinity. In order to do that we will
use the central Poincar\'e projection method. Thus, to obtain the
critical points at infinity we introduce spherical coordinates
($\rho$ is the inverse of $r=\sqrt{x_\phi^2+x_\vphi^2+y^2},$ then,
$\rho\rightarrow 0$ as $r\rightarrow \infty$):

\begin{align}
& x_\phi=\frac{1}{\rho}\sin\theta_1\cos\theta_2,\\
& y=\frac{1}{\rho}\sin\theta_1\sin\theta_2,\\
& x_\varphi=\frac{1}{\rho}\cos\theta_1
\end{align} where $0\leq \theta_1\leq \pi$ and $-\pi<\theta_2\leq
\pi,$ and $0<\rho<\infty.$

Defining the time derivative $f'\equiv \rho\dot f,$ the system
(\ref{eqxphi}-\ref{eqy}), can be written as

\be \rho'=\frac{1}{2} \left(\cos ^2\theta_1-\cos (2\theta_2) \sin
^2\theta_1\right)+2 n \cos\theta_1 \sin ^2\theta_1 \sin^2\theta_2
\rho +O\left(\rho ^2\right)\label{radial}. \ee

and

\begin{align}
&\theta_1'=n \cos (2\theta_1) \sin \theta_1 \sin ^2\theta_2-\cos
\theta_1 \sin\theta_1 \sin
   ^2\theta_2 \rho +O\left(\rho ^2\right),\nonumber\\
   &\theta_2'=(n \cos \theta_1 \cos \theta_2+
   m \sin \theta_1) \sin \theta_2-\cos \theta_2 \sin \theta_2 \rho
   +O\left(\rho ^2\right).\label{eqangular}
\end{align}

\begin{table*}[t]
\caption[crit]{Stability of the critical points at infinity. We
use the notation $\delta=m^2-n^2$ and $\lambda^\pm=n\cos
\theta_1^\star\pm m\sin\theta_1^\star.$}
\label{critatinfinityprop}
\begin{center}
\begin{tabular}{@{\hspace{4pt}}c@{\hspace{14pt}}c@{\hspace{18pt}}c@{\hspace{18pt}}c@{\hspace{2pt}}}
\hline
\hline\\[-0.3cm]
Name & $(\lambda_1,\lambda_2)$ &$\rho'$&Stability\\[0.1cm]
\hline\\[-0.2cm]
$P_1^\pm$&  $(-n,n)$ & $>0$ & saddle\\[0.2cm]
$P_2^\pm$ & $(-n,n)$ &
$>0$  & saddle \\[0.2cm]
$P_3^\pm$&$\left(\frac{\sqrt{2}\delta}{n},\frac{\delta}{\sqrt{2}n}\right)$&$\left\{\begin{array}{cc}
  >0, &  \delta<0 \\
  <0, & \delta>0 \\
\end{array}\right.$&source if $n<0, n<m<-n$\\[0.2cm]
\quad & \quad& \quad & saddle otherwise\\[0.2cm]
$P_4^\pm$&$\left(-\frac{\sqrt{2}\delta}{n},-\frac{\delta}{\sqrt{2}n}\right)$
&$\left\{\begin{array}{cc}
  >0, &  \delta<0 \\
  <0, & \delta>0 \\
\end{array}\right.$& source if $n>0, -n<m<n$\\[0.2cm]
\quad  & \quad& \quad & saddle otherwise\\[0.2cm]
$P_5$& $\left(0,\lambda^+\right)$& $\left\{\begin{array}{cc}
  <0, &  \frac{\pi}{4}<\theta_1^\star<\frac{3\pi}{4} \\
  >0, & \text{otherwise} \\
\end{array}\right.$&nonhyperbolic\\[0.2cm]
$P_6$& $\left(0,\lambda^-\right)$&$\left\{\begin{array}{cc}
  <0, &  \frac{\pi}{4}<\theta_1^\star<\frac{3\pi}{4} \\
  >0, & \text{otherwise} \\
\end{array}\right.$&nonhyperbolic
\\[0.4cm]
\hline \hline
\end{tabular}
\end{center}
\end{table*}

Since equation (\ref{radial}) does not depends of the radial
component at the limit $\rho\rightarrow 0,$ we can obtain the
critical points at infinity by solving equations (\ref{eqangular})
in the limit $\rho\rightarrow 0.$ Thus, the critical points at
infinite must satisfy the compatibility conditions

\bea & \cos (2 \theta_1) \sin
   \theta_1 \sin ^2\theta_2=0,\nonumber\\
   & (n \cos \theta_1 \cos
   \theta_2+m \sin \theta_1) \sin \theta_2=0.
   \label{compatibility}
\eea

First, we examine the stability of the pairs
$(\theta_1^\star,\theta_2^\star)$ satisfying the compatibility
conditions (\ref{compatibility}) in the plane
$\theta_1$-$\theta_2$, and then, we examine the global stability
by substituting in (\ref{radial}) and analyzing the sign of
$\rho'(\theta_2^\star,\theta_2^\star).$ In table
\ref{critatinfinity} it is offered information about the location
and existence conditions of these critical points. In table
\ref{critatinfinityprop} we summarize the stability properties of
these critical points.

Let us describe the cosmological solutions associated with the
critical points at infinity.

The cosmological solutions associated to the critical points
$P_1^\pm$ and $P_2^\pm$ have the evolution rates
$\dot\phi^2/V=0,\, \dot\phi/\dot\varphi=0$ and
$H/\dot\varphi\equiv\rho/\sqrt{6}\rightarrow 0.$ \footnote{Do not
confuse $\rho$ with the matter energy density, the latter denoted
by $\rho_{\rm M}.$} These solutions are always saddle points at
infinity. The critical points $P_3^\pm$ and $P_4^\pm$ are sources
provided $n<0, n<m<-n$ or $n>0, -n<m<n,$ respectively. They are
saddle points otherwise. The associated cosmological solutions to
$P_3^\pm$ have the evolution rates
$\dot\phi^2/V=\frac{2m^2}{n^2-m^2},\, \dot\phi/\dot\varphi=-m/n,$
and $H/\dot\phi\equiv -n\rho/(\sqrt{3} m)\rightarrow 0,$ and
$H/\dot\varphi\equiv \rho/\sqrt{3}\rightarrow 0,$ whereas the
associated cosmological solutions to $P_4^\pm$ have the evolution
rates $\dot\phi^2/V=\frac{2m^2}{n^2-m^2},\,
\dot\phi/\dot\varphi=-m/n,$ and $H/\dot\phi\equiv n\rho/(\sqrt{3}
m)\rightarrow 0,$ and $H/\dot\varphi\equiv
-\rho/\sqrt{3}\rightarrow 0.$ The curves of critical points $P_5$
and $P_6$ are nonhyperbolic. The associated cosmological solutions
have expansion rates (valid for $\theta_1^\star\neq\pi/4$)
$V/\dot\phi^2=0, \dot\phi/\dot\varphi=\tan \theta_1^\star,
H/\dot\varphi=\rho \sec\theta_1^\star/\sqrt{6}\rightarrow 0,$ and
$V/\dot\phi^2=0, \dot\phi/\dot\varphi=-\tan \theta_1^\star,
H/\dot\varphi=\rho \sec\theta_1^\star/\sqrt{6}\rightarrow 0,$
respectively.

\subsection{Models with negative curvature}\label{sectionIV}

In this section we investigate negative curvature models.

\subsubsection{Normalization, state space and dynamical
system.}\label{sectionIVA}

For the investigation of negative curvature models we shall use
the normalized variables: $(x_\phi,\,x_\varphi,\,y,\,\Omega)$,
defined by \begin{eqnarray} x_\phi=\frac{\dot\phi}{\sqrt{6} H},\;
x_\varphi=\frac{\dot\varphi}{\sqrt{6} H},  y=\frac{\sqrt
V}{\sqrt{3}H},\,\Omega= \frac{\rho_{\rm M}}{3 H^2}.\label{vars}
\end{eqnarray} This choice allows to recast the Friedmann equation (\ref{Friedmann1}) as \begin{eqnarray}
&&1-\left(x_\phi^2-x_\varphi^2+y^2+\Omega\right)= \Omega_k\geq
0,\label{ct} \end{eqnarray} where \be \Omega_k=-\frac{k}{a^2
H^2},\; k=-1,\,0. \ee Thus,
\begin{equation} 0\leq
x_\phi^2-x_\varphi^2+y^2+\Omega\leq 1 \label{constraint}.
\end{equation}

Let us introduce the new time variable, $\tau,$ such that
$\tau\rightarrow -\infty$ as $t\rightarrow 0$ and $\tau\rightarrow
+\infty$ as $t\rightarrow +\infty$. Since the time direction must
be preserved we can choose $d\tau=3 \epsilon H dt$ where
$\epsilon=\pm 1=\text{sign} (H).$

The field equations (\ref{vars}) are \begin{eqnarray}
&& x_\phi'=\epsilon\left(\sfrac13\,\left(q-2\right)\,x_\phi+m y^2\right),\nonumber\\
&& x_\varphi'=\epsilon\left(\sfrac13\,\left(q-2\right)\,x_\varphi-n y^2\right),\nonumber\\
&& y'=\epsilon\left( \sfrac{1}{3}(1+q) - m\,x_\phi -
n\,x_{\varphi}
      \right) \,y, \nonumber\\
&& \Omega'=\sfrac{1}{3}\epsilon\left(2\,q- 1 \right)
\,\Omega.\label{PhaseSpaceeqs}
\end{eqnarray}

Where $q=2\left(x_\phi^2-x_\varphi^2\right)
-y^2+\sfrac{1}{2}\Omega,$ is the expression for the deceleration
parameter. The DE EoS parameter, $w,$ can be rewritten, in terms
of the phase variables, as
\begin{equation}
w =\frac{x_\phi^2-x_\varphi^2-y^2}{x_\phi^2-x_\varphi^2+y^2}.
\end{equation}

Notice that the evolution equation (\ref{PhaseSpaceeqs} c) is form
invariant under the coordinate transformation $y\rightarrow
\epsilon y.$  Then, the sign of $\epsilon y$ is invariant by
proposition 4.1 in \cite{REZA}, in such way that we can assume,
without lost generality, for fixed $\epsilon$, $\epsilon y\geq 0.$
Hence, for each choice of sign of $\epsilon,$ the equations
(\ref{PhaseSpaceeqs}) define a flow in the phase space
\begin{eqnarray} &\Psi^\pm=\{\left(x_\phi, x_\varphi,
y,\Omega\right): 0\leq x_\phi^2-x_\varphi^2+y^2+\Omega\leq 1,
\nonumber\\& x_\phi^2-x_\varphi^2+y^2\geq 0,\Omega\geq 0,\epsilon
y\geq 0\}.\label{space}
\end{eqnarray}

\subsubsection{Form invariance under coordinate
trasformations.}\label{sectionIVC}

First recall that the positive ``branch'' ($\epsilon=+1$) describe
the dynamics of models ever expanding and the negative ``branch''
($\epsilon=-1$) describes the dynamics for contracting models. The
system is form invariant under the change $\epsilon\rightarrow
-\epsilon,$ i.e., the system is symmetric under time-reversing. In
this way it is enough to characterize de dynamics in $\Psi^+.$

\begin{table*}[t]\caption[Critical points of the system \ref{PhaseSpaceeqs}.]
{Coordinates and existence conditions for the critical points of
the system \ref{PhaseSpaceeqs}. We have used the notation
$\delta={m}^2-{n}^2.$  The subindexes in the labels have the
following meaning: the left subindex (denoted by $\epsilon=\pm 1$)
indicates when the model is expanding ($+$)or contracting ($-$);
the right subindex denotes the sign of $x_\phi$ (i.e., the sign of
$\dot \phi$) and it is displayed by the sign
$\pm.$}\label{table2a}
\begin{center}
\begin{tabular}{@{\hspace{4pt}}c@{\hspace{10pt}}c@{\hspace{10pt}}c@{\hspace{2pt}}}
\hline
\hline\\[-0.3cm]
Label & Coordinates:& Existence \\
      & $(x_\phi,x_\varphi,y,\Omega)$ & \\[0.1cm]
\hline\\[-0.3cm]
${}_{\pm}K_{\pm}$ & $(\pm\sqrt{1+{x_\varphi^\star}^2},x_\varphi^\star,0,0)$ & All $m$ and $n$ \\[0.2cm]
${}_{\pm}M$ & $(0,0,0,0)$ & All $m$ and $n$ \\[0.2cm]
${}_{\pm}F$ & $(0,0,0,1)$ & All $m$ and $n$  \\[0.4cm]
${}_\pm SF$ & $(m,-n,\epsilon\sqrt{1-\delta},0)$ & $\delta<1$  \\[0.2cm]
${}_\pm CS$  & $(\frac{m }{3\delta},-\frac{n }{3\delta},\frac{\epsilon\sqrt{2}}{3\sqrt{\delta}},0)$ & $\delta>\frac{1}{3}$ \\[0.2cm]
${}_{\pm}MS$ &
$(\frac{m}{2\delta},-\frac{n}{2\delta},\frac{\epsilon\sqrt{1}}{2\sqrt{\delta}},\sqrt{1-\frac{1}{2\delta}})$
& $\delta>\frac{1}{2}$  \\[0.4cm]
\hline \hline
\end{tabular}
\end{center}
\end{table*}

\begin{table*}[t]\caption[Properties of the cosmological solutions associated to the critical points
of the system \ref{PhaseSpaceeqs}.]{DE EoS parameter ($w$),
deceleration parameter ($q$), fractional energy densities, and
eigenvalues of the perturbation matrix associated to the critical
points of the system \ref{PhaseSpaceeqs}. We use the notation
$\lambda^\pm=n x_\varphi^\star\pm
m\sqrt{1+{x_\varphi^\star}^2}.$${}_{\pm}{K}_{\pm},$ ${}_{\pm}{F},$
${}_\pm {SF}$ ${}_{\pm}{MS}$ corresponds to $k=0$, the eigenvalues
of these points in the invariant set of zero-curvature models are
the same as displayed in the table but the first from the
left.}\label{table2b}
\begin{center}
\begin{tabular}{@{\hspace{4pt}}c@{\hspace{10pt}}c@{\hspace{10pt}}c@{\hspace{8pt}}c@{\hspace{8pt}}c@{\hspace{2pt}}}
\hline
\hline\\[-0.3cm]
Label & $w$& $q$&  $\Omega_m,\Omega_{de},\Omega_k$ & Eigenvalues \\
     \\[0.1cm]
\hline\\[-0.3cm]
${}_{\pm}K_{\pm}$ &  $1$ & $2$ & $0,1,0$  & $\frac{4}{3}\epsilon,0,\epsilon\left(1-\lambda^\pm\right),\epsilon$\\[0.2cm]
${}_{\pm}M$  & - & $0$ & $0,0,1$ & $-\frac{2}{3}\epsilon,-\frac{2}{3}\epsilon,\frac{1}{3}\epsilon,-\frac{1}{3}\epsilon$ \\[0.2cm]
${}_{\pm}F$ & - & $\frac{1}{2}$ & $1,0,0$ & $\frac{1}{2}\epsilon,-\frac{1}{2}\epsilon,-\frac{1}{2}\epsilon,\frac{1}{3}\epsilon$\\[0.4cm]
${}_\pm SF$ & $-1+2\delta$ & $-1+3\delta$ & $0,1,0$ & $2\left(\delta-\frac{1}{3}\right)\epsilon,\left(\delta-1\right)\epsilon,\left(\delta-1\right)\epsilon,\left(2\delta-1\right)\epsilon$  \\[0.2cm]
${}_\pm CS$ & $-\frac{1}{3}$ & $0$& $0,\frac{1}{3\delta},1-\frac{1}{3\delta}$ & $-\frac{2}{3}\epsilon,-\frac{1}{3}\epsilon,-\frac{1}{3}\left(\epsilon\pm\sqrt{\frac{4}{3\delta}-3}\right)$ \\[0.2cm]
${}_{\pm}MS$ &  $0$ & $\frac{1}{2}$  &
$1-\frac{1}{2\delta},\frac{1}{2\delta},0$ &
$\frac{1}{3}\epsilon,-\frac{1}{2}\epsilon,
-\frac{1}{4}\left(\epsilon\pm
\sqrt{\left(-7+\frac{4}{\delta}\right)}\right)$\\[0.4cm]
\hline \hline
\end{tabular}
\end{center}
\end{table*}

\subsubsection{Monotonic functions.}\label{sectionIVD}

Let be defined in the phase space $\Psi^+$ (or $\Psi^-$, depending
of the choice of $\epsilon$) the function

\begin{equation}
M=\frac{{\left( n\,x_\phi + m\,x_\varphi \right) }^2\,{\Omega }^2}
    {{\left( 1 - {x_\phi}^2 + {x_\varphi}^2 -
          y^2 - \Omega  \right) }^3},\; M'=-2\epsilon M. \label{monotonic1}
\end{equation}
This is a monotonic function for $\Omega>0$ and $n\,x_\phi +
m\,x_\varphi\neq 0.$ Then, the existences of such monotonic
function rule out periodic orbits, recurrent orbits, or homoclinic
orbits in the phase space and also, there is possible global
results from the local stability analysis of critical points.
Additionally, from the expresion of $M$ one can see inmediatly
that $\Omega\rightarrow 0,$ or $n\,x_\phi + m\,x_\varphi
\rightarrow 0$ o $\left|n\,x_\phi + m\,x_\varphi\right|
\rightarrow +\infty$ (implying $x_\phi $ or $x_\varphi$ or both
diverge) or $\Omega_k\rightarrow 0$ asymptotically.

\subsubsection{Local analysis of critical points.}\label{sectionIVE}

By the discussion about the invariance of the system, it is
sufficient characterize dynamically the critical points ${}_+
K_\pm,$ ${}_+ M,$ ${}_+ F$ ${}_+ SF$ ${}_+ CS$ y ${}_+ MS,$ in the
phase space $\Psi^+.$ In tables \ref{table2a} and \ref{table2b},
it is offered information about the location, existence and
eigenvalues of the critical points of the system
(\ref{PhaseSpaceeqs}) in the phase space (\ref{space}) (for each
choice of $\epsilon$) and also, it is displayed the values of some
cosmological parameters associated to the corresponding
cosmological solutions.

Now we shall investigate the local stability of the critical
points (and curves of critical points). We shall characterize de
associated cosmological solutions.

The set of critical points ${}_{\pm}K_{\pm}$ and the isolated
critical points ${}_{\pm}M$ are located in the invariant set of
massless scalar field (MSF) cosmologies without matter. The
isolated critical points ${}_{\pm}F$ are located in the invariant
set of MSF cosmologies with matter.

The arcs of hyperbolae ${}_{\pm}K_{\pm}$ parameterized by the real
value $x_\varphi^\star$ denote cosmological models dominated by
the energy density of DE ($\Omega_{de}\rightarrow 1$),
particularly by its kinetic energy. DE mimics a stiff fluid
solution. Since this are a set of critical points, then
necessarily, they have a zero eigenvalue. They are local sources
(and in general they constitute the past attractor in the phase
space $\Psi^+$) provided $n x_\varphi^\star\pm
m\sqrt{1+{x_\varphi^\star}^2}<1.$

The isolated critical points ${}_{\pm}M$ denote the Milne's
universe. They are non-hyperbolic. The critical points ${}_{\pm}F$
represent flat FRW solutions (dominated by matter). They are
hyperbolic. For this points the quintom field vanishes, then, the
DE's cosmological parameters are not applicable to this points.

The stable manifold of ${}_+M$ is 3-dimensional and it is tangent
at the point to the 3-dimensional space
$(x_\phi,x_\varphi,\Omega)$ whereas the unstable one is
1-dimensional and tangent to the axis $y.$ This means the the
critical point ${}_+M$ is unstable to perturbations in $y$. The
critical point ${}_+F$ have a 2-dimensional stable manifold
tangent at the point to the plane $(x_\phi,x_\varphi)$ and a
2-dimensional unstable manifold tangent at the critical point to
the plane $(y,\Omega).$

The isolated critical points ${}_\pm SF$ and ${}_\pm CS$ denotes
cosmological solutions dominated by quintom dark energy and
curvature scaling solutions, respectively. These are located in
the invariant set of MSF cosmologies without matter ($\Omega=0$).
The critical points ${}_\pm MS$ (belonging to the invariant set of
MSF cosmologies with matter ($\Omega>0$)) represent flat matter
scaling solutions.

The stable manifold of ${}_+ SF$ in $\Psi^+$ is 4-dimensional
provided $\delta<1/3.$ In this case ${}_+ SF$ is the global
attractor on $\Psi^+$. ${}_+ SF$ is a saddle with a 3-dimensional
stable manifold, if $\frac{1}{3}<\delta<\frac{1}{2}$ or
2-dimensional if  $\frac{1}{2}<\delta<1.$

The isolated critical points ${}_\pm CS$ are non-hyperbolic if
$\delta=\frac{1}{3}.$ On the other hand, the critical points
${}_\pm MS$ are non-hyperbolic if $\delta=\frac{1}{2}.$

${}_+CS$ is stable (with a 4-dimensional stable manifold) and
then, it is a global attractor provided
$\frac{1}{3}<\delta\leq\frac{4}{9}$ (in this case all the
eigenvalues are real) or if $\delta>\frac{4}{9}$ (in which case
there exists two complex conjugated eigenvalues in such way that
the orbits initially at the subspace spanned by the corresponding
eigenvectors spiraling toward the critical point).

Let us notice that ${}_+MS$ is the global attractor of the system
(it have a 4-dimensional stable manifold) only if
$0<\gamma<\sfrac{2}{3},\,\delta>\sfrac{\gamma}{2}$ (where $\gamma$
denotes the barotropic index of the perfect fluid). Since we are
assuming $\gamma=1$ (i.e., dust background) then, the critical
point ${}_+MS$ is a saddle. It have a 3-dimensional stable
manifold if $\frac{1}{2}<\delta\leq\frac{4}{7}$ (in which case all
the eigenvalues are real) or if  $\delta>\frac{4}{7}$ (in which
case there are two complex conjugated eigenvalues and then the
orbits initially at the subspace spanned by the corresponding
eigenvalues spiral in towards the critical point).

\subsubsection{Bifurcations.}\label{sectionIVF}

Observe that the critical points ${}_\pm MS$ and ${}_\pm SF$ are
the same as $\delta\rightarrow \frac{1}{2}^+$.  ${}_+ SF$ (${}_-
SF$) coincide with a point in the arc ${}_+K_+$ (${}_-K_-$) as
$\delta\rightarrow 1^-.$ This values of $\delta$ where the
critical points coincide correspond to bifurcations since the
stability changes.

\subsubsection{Typical behavior.}\label{sectionIVG}

Once the attractors have been identified one can give a
quantitative description of the physical behaviour of a typical
open ($k=-1$) quintom cosmology. For example, for ever expanding
cosmologies, near the initial singularity the model behave as de
flat FRW with stiff fluid (DE mimics a stiff fluid) represented by
a critical point in ${}_+K_+$ or in  ${}_+K_-,$ depending on the
selection of the free parameters $m, n$ and $x_\varphi^\star$ (see
table \ref{table4}). Whenever ${}_+CS$ exists (i.e., provided
$\delta>\sfrac{1}{3}$) it is the global attractor of the system.
In absence of this type of points, i.e., if $\delta<\sfrac{1}{3}$,
the late time dynamics is determined by the critical point
${}_+SF$, i.e., the universe will be accelerated, almost flat
($\Omega_k\rightarrow 0$) and dominated by DE
($\Omega_{de}\rightarrow 1$).  DE behaves like quintessence ($-1 <
q < 0$, i.e., $-1 < w < -\sfrac{1}{3}$) or a phantom field ($q <
-1$, i.e., $w < -1$) if $\delta> 0$ or $\delta< 0,$ respectively.
This means that, typically, the ever expanding open quintom model
crosses the phantom divide (DE EoS parameter have values less than
$-1$). \footnote{For flat models, is well known the, whenever it
exists, (i.e., provided $\delta>\frac{1}{2}$) the attractor is
${}_+MS$ (denoted by $T$ in \cite{Lazkoz:2006pa}). When we include
curvature, the stability of the matter scaling solution is
transferred to the curvature scaling solution, as we prove here.}
The intermediate dynamics will be governed by the critical points
${}_+CS,$ ${}_+MS,$ y ${}_+M,$ which have the highest
lower-dimensional stable manifold.

\begin{table*}[ht!]
\caption[Summary of attractors of the system \ref{PhaseSpaceeqs}.]{Summary of attractors of the system \ref{PhaseSpaceeqs}. Observe that, whenever exists, the solution dominated by curvature ${}_-CS$ (${}_+CS$)
is the past (the future) attractor for $\epsilon=-1$, i. e., for
contracting models ($\epsilon=1$, i.e., for expanding models). We
use the notation $\lambda^\pm=n x_\varphi^\star\pm
m\sqrt{1+{x_\varphi^\star}^2}.$}\label{table4}
\begin{center}
\begin{tabular}{@{\hspace{4pt}}c|@{\hspace{10pt}}c@{\hspace{10pt}}c@{\hspace{2pt}}}
\hline
\hline\\[-0.3cm]
Restrictions &  Past attractor   & Future attractor \\[0.1cm]
\hline\\[-0.3cm]
$\epsilon=-1$ & $\begin{array}{cc}
{}_-SF & \text{if}\;\delta<\sfrac{1}{3}\\[0.2cm]
{}_-CS & \text{if}\;\delta>\sfrac{1}{3}\\[0.2cm]
\end{array}$   & ${}_-K_\pm\,\text{if}\;\,\lambda^\pm>-1$  \\[0.2cm]\hline\\[0.1cm]
$\epsilon=1$ & ${}_+K_\pm\,\text{if}\;\lambda^\pm<1$  &
$\begin{array}{cc}
{}_+SF & \text{if}\;\delta<\sfrac{1}{3}\\[0.2cm]
{}_+CS & \text{if}\;\delta>\sfrac{1}{3}\\[0.2cm]
\end{array}$ \\[0.4cm]\hline \hline
\end{tabular}
\end{center}
\end{table*}

For contracting models, the typical behavior, is in some way, the
reverse of the above. If $\delta<\sfrac{1}{3}$ the early time
dynamics is dominated by ${}_-CS.$ Otherwise, if
$\delta>\sfrac{1}{3},$ the past attractor is ${}_-SF$, i.e., the
model is accelerating, close to flatness ($\Omega_k\rightarrow 0$)
and dominated by DE. The intermediate dynamics is dominated at
large extent by the critical points ${}_-CS,$ ${}_-MS,$ y ${}_-F,$
which have the highest lower-dimensional stable manifold. A
typical model behaves at late times as a flat FRW universe with
stiff fluid (i.e., ME mimics a stiff fluid) represented by the
invariant sets ${}_-K_+$ or ${}_-K_-$, depending on the choice of
the  values of the free parameters $m, n$ y $x_\varphi^\star.$

\subsection{Models with positive curvature}\label{sectionV}

In this section we investigate positive curvature models we shall
make use of the variables similar but not equal to those defined
in \cite{Coley:2003mj} section VI.A.

\subsubsection{Normalization, state space and dynamical system.}\label{sectionVA}

Let us introduce the normalization factor
\begin{equation}
\hat{D}= 3\sqrt{H^2+a^{-2}}.
\end{equation}

Observe that $$\hat{D}\rightarrow 0\Leftrightarrow H\rightarrow
0,\, a\rightarrow +\infty$$ (i.e., at a singularity). This means
that it is not possible that $\hat{D}$ vanishes at a finite time.

Let us introduce the following normalized variables
$(Q_0,\,\hat{x}_\phi,\,\hat{x}_\varphi,\,\hat{y},\,\hat{\Omega})$,
given by
\begin{equation}
Q_0= \frac{3 H}{\hat{D}},\;
\hat{x}_\phi=\sqrt{\frac{3}{2}}\frac{\dot\phi}{\hat{D}},
\hat{x}_\vphi=\sqrt{\frac{3}{2}}\frac{\dot\varphi}{\hat{D}},\hat{y}=\frac{\sqrt
3 V}{\hat{D}},\,\hat{\Omega}= \frac{3 \rho_{\rm
M}}{\hat{D}^2}.\label{vars2}
\end{equation}

From the Friedmann equation we find
\begin{equation}
0\leq \hat{x}_\phi^2-\hat{x}_\phi^2+\hat{y}^2=1-\hat{\Omega}\leq
1\label{Friedmann}
\end{equation}
and by definition
\begin{equation}
-1\leq Q_0\leq 1. \label{Q0}
\end{equation}

By the restrictions (\ref{Friedmann}, \ref{Q0}), the state
variables are in the state space
\begin{equation}
 \hat{\Psi}=\{\left(Q_0,\hat{x}_\phi,\hat{x}_\varphi,\hat{y}\right): 0\leq \hat{x}_\phi^2-\hat{x}_\phi^2+\hat{y}^2\leq 1, -1\leq Q_0\leq 1\}.\label{statespace2}
\end{equation}
As before, this state space is not compact.

Let us introduce the time coordinate $$' \equiv
\frac{d}{d\hat{\tau}}=\frac{3}{\hat{D}}\frac{d}{d t}.$$ $\hat{D}$
has the evolution equation
$$\hat{D}'=-3 Q_0\hat{D}\left(\hat{x}_\phi^2-\hat{x}_\varphi^2+\frac{1}{2}\hat{\Omega}\right)$$ where
$$\hat{\Omega}=1-\left(\hat{x}_\phi^2-\hat{x}_\varphi^2+\hat{y}\right).$$
This equation decouples from the other evolution equations. Thus, a reduced set of evolution equations is obtained.
\begin{eqnarray}
&& Q_0'= {\left( 1 - Q_0^2 \right) \,\left( 1 -
      3\,\Xi\right)},\nonumber\\
&& \hat{x}_\phi'=3\,m\,\hat{y}^2 + 3\,Q_0\,\hat{x}_\phi\,\left( -1 + \Xi \right),\nonumber \\
&& \hat{x}_\varphi'=-3\,n\,\hat{y}^2 + 3\,Q_0\,\hat{x}_\varphi\,\left( -1 + \Xi \right),\nonumber\\
&&
\hat{y}'=-3\,\hat{y}\left(m\,\hat{x}_\phi+n\,\hat{x}_\varphi-Q_0\,\Xi
\right).\label{ds2}
\end{eqnarray}
Where $\Xi={\hat{x}_\phi}^2 - {\hat{x}_\varphi}^2 +
\frac{1}{2}\,\hat{\Omega}.$

There is also an auxiliary evolution equation
\begin{equation}
\hat{\Omega}'=-Q_0\,\left( -2\,\left( {\hat{x}_\phi}^2 -
{\hat{x}_\varphi}^2 \right)  +  \,\left( 1 - \hat{\Omega}  \right)
\right) \, \hat{\Omega}. \label{eqOmega}
\end{equation}

It is useful to express some cosmological parameters in terms of
our state variables. \footnote{We have defined $\Omega_k\equiv
\frac{k}{a^2 H^2}=\frac{1}{a^2 H^2}.$}
$$\left(\Omega_m,\Omega_{de},\Omega_k,q\right)=\left(\hat{\Omega},1-\hat{\Omega},Q_0^2-1,-1+3\Xi\right)/Q_0^2,$$ and $$w=\frac{\hat{x}_\phi^2-\hat{x}_\varphi^2-\hat{y}^2}{\hat{x}_\phi^2-\hat{x}_\varphi^2+\hat{y}^2}.$$

\subsubsection{Invariance under coordinate transformations.}\label{sectionVC}

Observe that the system (\ref{ds2}, \ref{eqOmega}) is invariant
under the transformation of coordinates

\begin{equation}
\left(\hat{\tau}, Q_0, \hat{x}_\phi,
\hat{x}_\varphi,\hat{y},\hat{\Omega}\right)\rightarrow
\left(-\hat{\tau}, -Q_0, -\hat{x}_\phi, -\hat{x}_\varphi,
\hat{y},\hat{\Omega}\right)\label{discrete2}.
\end{equation}
Thus, it is sufficient to discuss the behaviour in one part of the
phase space, the dynamics in the other part being obtained via the
transformation (\ref{discrete2}). In relation with the possible
attractors of the system we will characterize those corresponding
to the ``positive'' branch. The dynamical behavior of the critical
points in the ``negative'' branch is determined by the
transformation (\ref{discrete2}).

\subsubsection{Monotonic functions.}\label{sectionVD}

The function \begin{equation}
N=\frac{\left(n\,\hat{x}_\phi+m\,\hat{x}_\varphi\right)^2\,\hat{\Omega}^2}{\left(1-Q_0^2\right)^{3}},\;
N'=-6\,Q_0\,N\label{monotonic2}
\end{equation} is monotonic in the regions $Q_0<0$ and $Q_0>0$ for $Q_0^2\neq 1,\, n\,\hat{x}_\phi+n\,\hat{x}_\varphi\neq 0,\, \hat{\Omega}>0.$ Hence, there can be no periodic orbits or recurrent orbits in the interior of the phase space.
Furthermore, it is possible to obtain global results. From the
expression $N$ we can immediately see that asymptotically
$Q_0^2\rightarrow 1$ or $n
\hat{x}_\phi+m\hat{x}_\varphi\rightarrow 0$ or
$\hat{\Omega}\rightarrow 0.$

\subsubsection{Local analysis of critical points.}\label{sectionVE}

In the tables \ref{table6} and \ref{table6b} it is summarized the
location, existence conditions, some properties of the critical
points and the eigenvalues of the linearized system around each
critical point.

\begin{table*}[ht!]\caption{Critical points of the system (\ref{ds2}).
We use the same notation as in table \ref{table2a}.}\label{table6}
\begin{center}
\begin{tabular}{@{\hspace{4pt}}c@{\hspace{14pt}}c@{\hspace{10pt}}c@{\hspace{2pt}}}
\hline
\hline\\[-0.3cm]
Label & Coordinates:& Existence \\
      & $(Q_0, \hat{x}_\phi,\hat{x}_\varphi,\hat{y})$ &\\[0.1cm]
\hline\\[-0.3cm]
${}_{\pm}\hat{K}_{\pm}$ & $(\epsilon,\pm\sqrt{1+{x_\varphi^\star}^2},x_\varphi^\star,0)$ & All $m$ and $n$ \\[0.2cm]
${}_{\pm}\hat{F}$ & $(\epsilon,0,0,0)$ & All $m$ and $n$  \\[0.2cm]
${}_\pm \hat{SF}$ & $(\epsilon,m\epsilon,-n\epsilon,\sqrt{1-\delta})$ & $\delta<1$ \\[0.2cm]
${}_\pm \hat{CS}$  & $(\sqrt{3\delta}\epsilon,\frac{m \epsilon}{\sqrt{3\delta}},-\frac{n \epsilon}{\sqrt{3\delta}},\sqrt{\frac{2}{3}})$ & $0<\delta<\frac{1}{3}$\\[0.2cm]
${}_{\pm}\hat{MS}$ & $(\epsilon,\frac{m}{2\delta},-\frac{n}{2\delta},\frac{\sqrt{1}}{2\sqrt{\delta}},\sqrt{1-\frac{1}{2\delta}})$ & $\delta>\frac{1}{2}$ \\[0.4cm]
\hline \hline
\end{tabular}
\end{center}
\end{table*}

\begin{table*}[t]\caption[Properties of the cosmological solutions associated to the critical points
of the system (\ref{ds2}).]{DE EoS parameter ($w$), deceleration
parameter ($q$), fractional energy densities, and eigenvalues of
the perturbation matrix associated to the critical points of the
system (\ref{ds2}). We use the notation $\lambda^\pm=n
x_\varphi^\star\pm m\sqrt{1+{x_\varphi^\star}^2}.$ When the flow
is restricted to the invariant sets $Q_0=\pm 1,$ the eigenvalues
associated to the critical points ${}_{\pm}\hat{F},$ ${}_\pm
\hat{SF}$ and ${}_\pm \hat{MS}$ and to the critical sets
${}_{\pm}\hat{K}_{\pm},$ are, in each case, the same as those
displayed, but the first from the left.}\label{table6b}
\begin{center}
\begin{tabular}{@{\hspace{4pt}}c@{\hspace{10pt}}c@{\hspace{10pt}}c@{\hspace{8pt}}c@{\hspace{8pt}}c@{\hspace{8pt}}c@{\hspace{2pt}}}
\hline
\hline\\[-0.3cm]
Label & $w$& $q$&  $\Omega_m,\Omega_{de},\Omega_k$ & Eigenvalues \\
     \\[0.1cm]
\hline\\[-0.3cm]
${}_{\pm}\hat{K}_{\pm}$ &  $1$ & $2$ & $0,1,0$  & $4\epsilon,0,3\left(\epsilon-\lambda^\pm\right),3\epsilon$ \\[0.2cm]
${}_{\pm}\hat{F}$ & - & $\frac{1}{2}$ & $1,0,0$ & $\epsilon,\frac{3}{2}\epsilon,-\frac{3}{2}\epsilon,-\frac{3}{2}\epsilon$\\[0.4cm]
${}_\pm \hat{SF}$ & $-1+2\delta$ & $-1+3\delta$ & $0,1,0$ & $2\left(3\delta-1\right)\epsilon,3\left(\delta-1\right)\epsilon,3\left(\delta-1\right)\epsilon,3\epsilon$  \\[0.2cm]
${}_\pm \hat{CS}$ & $-\frac{1}{3}$ & $0$& $0,\frac{1}{3\delta},1-\frac{1}{3\delta}$ & $-2\sqrt{3\delta}\epsilon,-\sqrt{3\delta}\epsilon\pm\sqrt{4-9\delta},-\sqrt{3\delta}\epsilon$  \\[0.2cm]
${}_{\pm}\hat{MS}$ &  $0$ & $\frac{1}{2}$  &
$1-\frac{1}{2\delta},\frac{1}{2\delta},0$  &
$\epsilon,-\frac{3}{2}\epsilon, -\frac{3}{4}\left(\epsilon\pm
\sqrt{\left(-7+\frac{4}{\delta}\right)}\right)\epsilon$ \\[0.4cm]
\hline \hline
\end{tabular}
\end{center}
\end{table*}

In the following we will characterize the dynamical behavior of
the cosmological solutions associated with them.

The critical points ${}_\pm\hat{K},{}_\pm\hat{F}, {}_\pm \hat{SF}$
and  ${}_\pm \hat{MS}$ represents flat FRW solutions.

\begin{figure}[!]
\begin{center}
\subfigure[normal][]{\includegraphics[width=12cm]{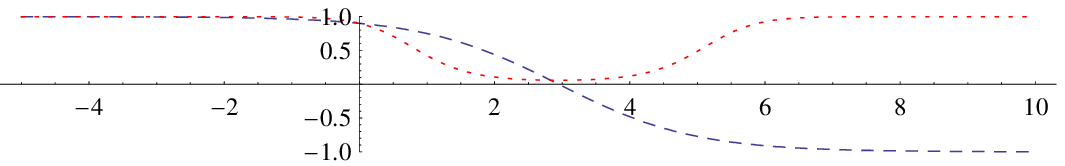}\label{Recollapsea}}
\subfigure[normal][]{\includegraphics[width=12cm]{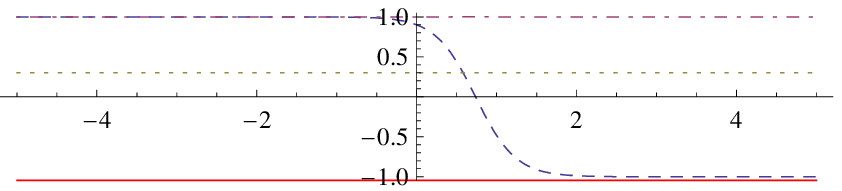}\label{Recollapseb}}
\caption{(Color online){\it The collapse of quintom cosmologies
with positive curvature for the values for the parameters
${m}=0.7,$ ${n}=0.3$ and $\gamma=1.$ In (a) we have selected the
initial conditions $Q_0(0)=\hat{\Omega}(0)=0.9,$ and
$\hat{x}_\phi(0)=\hat{x}_\varphi(0)=\hat{y}(0)=0.$ The dashed
(blue) line represents the evolution of $Q_0$ vs $\tau$ (observe
that $Q_0$ evolves from $1$ to $-1,$ and eventually takes zero
value). The dotted (red) line represents $\hat{\Omega}$ vs $\tau$.
This illustrates the existence of a closed FRW solution with no
scalar field starting from a big-bang at ${}_+\hat{F}$ and
recollapsing to a ``big-crunch'' at ${}_-\hat{F}.$ In (b) we have
selected the initial conditions $Q_0(0)=0.9,$ $\hat{\Omega}(0)=0$
and $\hat{x}_\phi(0)=-\sqrt{1+\hat{x}_\varphi(0)^2}$ with
$\hat{x}_\varphi(0)=0.3.$ The dashed (blue) line denotes $Q_0$ vs
$\tau.$ Observe that $Q_0$ goes from the value $1$ to $-1$ (i.e.,
the model collapses). The dot-dashed line denotes the evolution of
$\hat{x}_\phi^2-\hat{x}_\varphi^2$ vs $\tau$ (which is identically
equal to 1). The dotted (yellow) line denotes the value of
$\hat{x}_\varphi$ vs $\tau$ and the straight (red) line denotes
the value of $\hat{x}_\phi$ vs $\tau.$ This illustrates the
existence of orbits of the type ${}_+\hat{K}_-\rightarrow
{}_-\hat{K}_-.$ By choosing
$\hat{x}_\phi(0)=\sqrt{1+\hat{x}_\varphi(0)^2},$ with the same
initial conditions for the other variables, we obtain orbits of
the type ${}_+\hat{K}_+\rightarrow {}_-\hat{K}_+.$ These are
solutions starting from and recollapsing to a singularity (given
by a massless scalar field cosmology).}}
\end{center}\label{Recollapse}
\end{figure}

The set of critical  points ${}_+\hat{K}_\pm$  parameterised by
the real value $x_\varphi^\star$ represents stiff fluid
cosmological solutions (DE mimics a stiff fluid). It is the past
attractor for ever expanding models provided $n x_\varphi^\star\pm
m\sqrt{1+{x_\varphi^\star}^2}<1.$  As we proceed before, a simple
application of the symmetry (\ref{discrete2}), allows to the
identification of the future attractor for collapsing models: the
typical orbits tends asymptotically to ${}_-\hat{K}_\pm$  as
$\hat{\tau}\rightarrow \infty$ provided $n x_\varphi^\star\pm
m\sqrt{1+{x_\varphi^\star}^2}>-1,$ and $-1\leq Q_0<0.$ This fact
has interesting consequences. If $x_\varphi^\star$ is a fixed
value and $n$ and $m$ are such that $-1<n
x_\varphi^\star+m\sqrt{1+{x_\varphi^\star}^2}<1,$ then, there
exists one orbit of the type ${}_+\hat{K}_+\rightarrow
{}_-\hat{K}_+.$ If  $n$ and $m$ are such that $-1<n
x_\varphi^\star- m\sqrt{1+{x_\varphi^\star}^2}<1,$  then, there is
one orbit of the type ${}_+\hat{K}_-\rightarrow {}_-\hat{K}_-.$
These are solutions starting from and recollapsing to a
singularity given by a MSF cosmology (see figure
\ref{Recollapseb}).

The critical points ${}_\pm \hat{F}$ represent flat FRW solutions.
They hyperbolic. For these points the scalar fields vanish, so the
cosmological parameters associated to DE are not applicable to
these points. If $\delta>\sfrac{2}{3},$ the ustable (stable)
manifold of  ${}_+\hat{F}$ (${}_-\hat{F}$) is tangent to the
critical point and parallel to the plane $\hat{y}-Q_0.$ This means
that there is an orbit connecting ${}_+\hat{F}$ and ${}_-\hat{F}$
pointing towards ${}_-\hat{F}$ in the direction of the $Q_0$-axis.
It represents the closed FRW solution with no scalar field
starting from a big-bang at ${}_+\hat{F}$ and recollapsing to a
``big-crunch'' at ${}_-\hat{F}$ (see figure \ref{Recollapsea}).

The critical point ${}_+\hat{SF}$ represents a solution dominated
by the scalar field (with non-vanishing potential). It can be the
global attractor in the sets $0<Q_0<1$ or $Q_0=1$ (i.e., for ever
expanding models, or flat models) for the values of the parameters
displayed in \ref{table7}. It can be a phantom dominated solution
provided $\delta<0.$ It also can represent quintessence dominated
or de Sitter solutions.

The critical point ${}_+\hat{MS}$ exist if $\delta>\sfrac{1}{2}.$
They represent flat matter scaling solutions, for which both the
fluid and quintom are dynamically important. It is a saddle point.

For $0<\delta<\sfrac{1}{3}$ there exists the critical points
${}_\pm \hat{CS}$ for which the matter is unimportant, but
curvature is non-vanishing ($Q_0^2\neq 1$) and tracks the scalar
field. These are called curvature scaling solutions. The values of
its cosmological parameters are the same as for ${}_\pm {CS}$
(displayed in table \ref{table2b}), but it represents a different
cosmological solution with positive curvature. These critical
points are typically saddle points.

In table \ref{table7}, where we present a summary of attractors
for the quintom model with $k=1.$

\begin{table*}[ht!]\caption{Summary of attractors for the quintom model with
$k=1$(system (\ref{ds2})). We use the notation $\lambda^\pm=n
x_\varphi^\star\pm m\sqrt{1+{x_\varphi^\star}^2}.$}\label{table7}
\begin{center}
\begin{tabular}{@{\hspace{4pt}}c|@{\hspace{10pt}}c@{\hspace{10pt}}c@{\hspace{2pt}}}
\hline
\hline\\[-0.3cm]
Restrictions &  Past attractor    & Future attractor \\[0.1cm]
\hline\\[-0.3cm]
$Q_0=-1$ & $\begin{array}{cc}
{}_-\hat{SF} & \text{if}\;\delta<\sfrac{1}{2}\\[0.2cm]
{}_-\hat{MS}& \text{if}\;\delta>\sfrac{1}{2}\\[0.2cm]
\end{array}$   & ${}_-\hat{K}_\pm\,\text{if}\;\,\lambda^\pm>-1$  \\[0.2cm]\hline\\[0.1cm]
$-1<Q_0<0$ & $\begin{array}{cc}
{}_-\hat{SF} & \text{if}\;\delta<\sfrac{1}{3}\\[0.2cm]
\end{array}$   & as above \\[0.2cm]\hline\\[0.1cm]
$0<Q_0<1$ & ${}_+\hat{K}_\pm\,\text{if}\;\lambda^\pm<1$ &
$\begin{array}{cc}
{}_+\hat{SF} & \text{if}\;\delta<\sfrac{1}{3}\\[0.2cm]
\end{array}$  \\[0.2cm]\hline\\[0.1cm]
$Q_0=1$ & as above & $\begin{array}{cc}
{}_+\hat{SF} & \text{if}\;\delta<\sfrac{1}{2}\\[0.2cm]
{}_+\hat{MS}& \text{if}\;\delta>\sfrac{1}{2}\\[0.2cm]
\end{array}$ \\[0.4cm]
\hline \hline
\end{tabular}
\end{center}
\end{table*}

\subsubsection{Bifurcations.}\label{sectionVF}

Observe that the critical points ${}_\pm \hat{MS}$ y ${}_\pm
\hat{SF}$ coincides as $\delta\rightarrow \frac{1}{2}^+.$ ${}_\pm
\hat{CS}$ y ${}_\pm \hat{SF}$ coincides as
$\delta\rightarrow\frac{1}{3}^-.$ Additionally, ${}_+ \hat{SF}$
(${}_- \hat{SF}$) coincides with a point at the arc
${}_+\hat{K}_+$ (${}_-\hat{K}_-$) as $\delta\rightarrow 1^-.$ For
this values of $\delta$ a bifurcation occurs.

\subsubsection{Typical behaviour.}\label{sectionVG}

Once the attractors have been identified one can give a
quantitative description of the physical behaviour of a typical
closed quintom cosmology. For example, for ever expanding
cosmologies, near the big-bang a typical model behaves like a flat
FRW model with stiff fluid represented by the critical set
${}_+\hat{K}_+$ or by ${}_+\hat{K}_-$, depending on the choice of
the values of the free parameters $m, n$ and $x_\varphi^\star.$ If
$\delta<\sfrac{1}{3}$  and $0<Q_0<1$ the late time dynamics is
determined by ${}_+\hat{SF}$, (with the same physical properties
as ${}_+{SF}$). The intermediate dynamics will be governed to a
large extent by the fixed points ${}_+\hat{CS},$ ${}_+\hat{MS},$
and ${}_+\hat{F},$ which have the highest lower-dimensional stable
manifold. For flat models (i.e., in the invariant set $Q_0=1$),
the late time dynamics is determined by the critical point
${}_+\hat{SF}$ provided $\delta<\sfrac{1}{2}$ or ${}_+\hat{MS}$
provided $\delta>\sfrac{1}{2}.$

For contracting models, the typical behavior is, in one sense, the
reverse of the above. If $\delta<\sfrac{1}{3}$ and ${-1<Q_0<0}$
the early time dynamics is determined by ${}_-\hat{SF}.$ The
intermediate dynamics will be governed to a large extent by the
fixed points ${}_-\hat{CS},$ ${}_-\hat{MS},$ and ${}_-\hat{F},$
which have the highest lower-dimensional stable manifold. For flat
models (i.e., in the invariant set ${Q_0=-1}$), the early time
dynamics is determined by the critical point ${}_-\hat{SF}$ (or
${}_-\hat{MF}$) provided $\delta<\sfrac{1}{2}$
($\delta>\sfrac{1}{2}$). A typical model behaves at late times
like a flat FRW model with stiff fluid (i.e. the dark energy
mimics a stiff fluid) represented by the critical set
${}_-\hat{K}_+$ or by ${}_-\hat{K}_-$ depending on the choice of
the values of the free parameters $m, n$ and $x_\varphi^\star.$

\section{Observational Evidence for Quinstant Dark Energy Paradigm}

\subsection{The model}

Looking at the impressive amount of papers addressing the problem
of cosmic acceleration clearly shows that two leading candidates
to the dark energy throne are the old cosmological constant
$\Lambda$ and a scalar field $\phi$ evolving under the influence
of its self\,-\,interaction potential $V(\phi)$.

In the usual approach, one adds either a scalar field or a
cosmological constant term to the field equations. However, since
what we see is only the final effect of the dark energy
components, in principle nothing prevents us to add more than one
single component provided that the effective dark energy fluid
coming out is able to explain the data at hand. Moreover, as we
have hinted upon above, a single scalar field, while explaining
cosmic speed up, leads to a problematic eternal acceleration. A
possible way out of this problem has been proposed by some of us
\cite{RolLambda, Cardone2008, Nodal2008} through the introduction
of a negative cosmological term.

Motivated by those encouraging results, we therefore consider a spatially
flat universe filled by dust matter, radiation, scalar field and a
(negative) cosmological constant term. The Friedmann equations thus read\,:

\begin{equation}
H^2 = \frac{1}{3} \left [ \rho_M + \rho_r + \rho_{\Lambda} +
\frac{1}{2} \dot{\phi}^2 + V(\phi) \right ] \ ,
\label{eq: f1}
\end{equation}

\begin{equation}
2 \dot{H} + 3 H^2 = - \left [ \frac{1}{3} \rho_r - \rho_{\Lambda} +
\frac{1}{2} \dot{\phi}^2 - V(\phi) \right ] \ ,
\label{eq: f2}
\end{equation}
where we have used natural units with $8 \pi G = c = 1$.

\subsection{Matching with the data}
Notwithstanding how well motivated it is, a whatever model must be able to
reproduce what is observed. This is particularly true for the model we are
considering because the presence of a negative cosmological constant
introduces a positive pressure term potentially inhibiting the cosmic speed
up. Moreover, contrasting the model against the data offers also the
possibility to constrain its characteristic parameters and estimate other
derived interesting quantities, such as $q_0$, the transition redshift
$z_T$ and the age of the universe $t_0$. Motivated by these considerations,
we will therefore fit our model to the dataset described below
parametrizing the model itself with the matter density $\Omega_M$, the
scalar field quantities $(\Omega_{\phi}, w_0)$ and the dimensionless Hubble
constant $h$ (i.e., $H_0$ in units of $100 \ {\rm km/s/Mpc}$), while we
will set the radiation density parameter as $\Omega_r = 10^{-4.3}$ as in
\cite{FP04} from a median of different values reported in literature.

\subsubsection{The method and the data}

In order to constrain the model parameters we will consider several observational test:
 (a)the distance modulus $\mu= m - M$, i.e. the difference between the apparent and absolute
magnitude of an object at redshift $z$, (b) the gas mass fraction in galaxy clusters, (c)the measurement of the baryonic acoustic oscillation (BAO) peak in the large scale correlation function at $100 \ h^{-1} \ {\rm Mpc}$ separation
detected by Eisenstein et al. \cite{Eis05} using a sample of 46748 luminous
red galaxies (LRG) selected from the SDSS Main Sample \cite{SDSSMain}, (d) the shift parameter \cite{BET97} \footnote{a complete discusion about this observational test and the quinstant model can be found in \cite{Cardone2008}} and we maximize the following likelihood taking into account the above test:
\begin{equation}
{\cal{L}} \propto \exp{\left [ - \frac{\chi^2({\bf p})}{2} \right
]} \label{eq: deflike}
\end{equation}
where ${\bf p} = (\Omega_M, \Omega_{\phi}, w_0, h)$ denotes the set of
model parameters and the pseudo\,-\,$\chi^2$ merit function reads\,:

\begin{eqnarray}
\chi^2({\bf p}) & = & \sum_{i = 1}^{N}{\left [ \frac{\mu^{th}(z_i, {\bf p}) - \mu_i^{obs}}
{\sigma_i} \right ]^2} + \sum_{i = 1}^{N}{\left [
\frac{f_{gas}^{th}(z_i, {\bf p}) - f_{gas,i}^{obs}}{\sigma_i} \right ]^2}
\nonumber \\ ~ & + &
\displaystyle{\left [ \frac{{\cal{A}}({\bf p}) - 0.474}{0.017} \right ]^2} +
\displaystyle{\left [ \frac{{\cal{R}}({\bf p}) - 1.70}{0.03} \right
]^2}  + \left ( \frac{h - 0.72}{0.08} \right )^2 \ .
\label{eq: defchi}\
\end{eqnarray}

\subsubsection{Results}
Table 1 shows the best fit model parameters, median values and $1$ and $2\sigma$ ranges for
the parameters $(\Omega_{M}$, $\Omega_{\Lambda}$, $w_0$, $h$, $\Omega_{\phi})$.

\begin{table}[!ht]
\caption{Best fit ($bf$) and median ($med$) values and $1 \sigma$ and $2 \sigma$ ranges of the
parameters $(\Omega_M, \Omega_{\Lambda}, w_0, h,\Omega_{\phi}))$ as obtained from the likelihood analysis (from \cite{Cardone2008}).}

\begin{center}
\begin{tabular}{|c|c|c|c|c|}
\hline
Par & $bf$ & $med$ & $1 \sigma$ & $2 \sigma$ \\
\hline \hline
$\Omega_M$ & 0.283 & 0.307 & $(0.272, 0.352)$ & $(0.246, 0.410)$ \\
$\Omega_\Lambda$ & -0.072 & -0.298 & $(-0.54, -0.11)$ & $(-0.92, -0.02)$ \\
$w_0$ & -0.72 & -0.67 & $(-0.74, -0.60)$ & $(-0.79, -0.53)$ \\ $h$ & 0.632
& 0.620 & $(0.588, 0.654)$ & $(0.554, 0.692)$ \\ $\Omega_{\phi}$ & 0.789 &
0.989 & $(0.799, 1.226)$ & $(0.700, 1.574)$ \\
\hline
\end{tabular}
\end{center}
\end{table}

\begin{figure}[!]
\begin{center}
\subfigure[normal][]{\includegraphics[width=10cm]{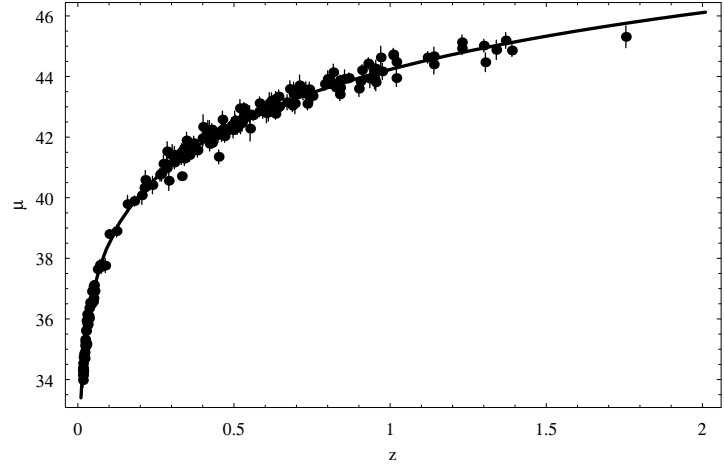}\label{fig:
sneiafit}}
\subfigure[normal][]{\includegraphics[width=10cm]{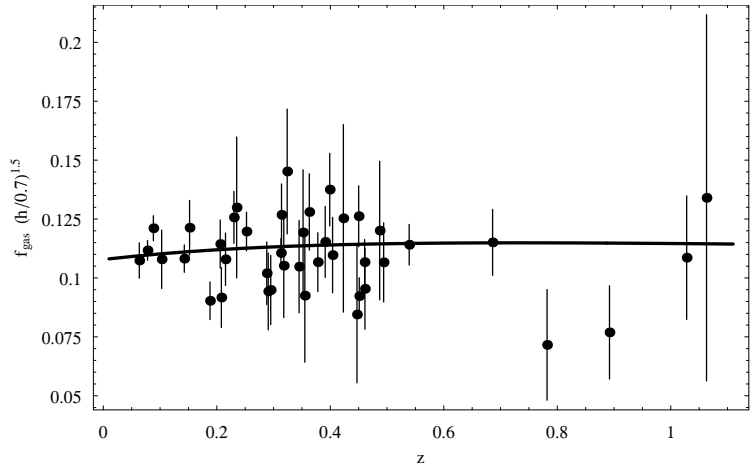}\label{fig:
gasfit}} \caption{(Color online) {\it (a) Best fit curve
superimposed to the data on the SNeIa Hubble diagram. (b) Best fit
curve superimposed to the data on the gas mass fraction. Note that
the theoretical curve plots indeed $f_{gas}(z) {\times}
(h/0.7)^{1.5}$ with $h$ set to its best fit value (from
\cite{Cardone2008}).}} \label{Bestfit}
\end{center}
\end{figure}

Figs.\ref{Bestfit} shows how well our best fit model reproduce the
data on the SNeIa Hubble diagram and gas mass fraction. The best
fit model is in quite good agreement with both the SNeIa and gas
data. Actually, the $\chi^2$ values are respectively $206$ and
$48$ to be contrasted with the number of datapoints, being $192$
and $42$ respectively. Besides the predicted values for the
acoustic peak and shift parameters are in satisfactory agreement
with the observed ones:
\begin{equation}
{\cal{A}} = 0.45 \ \ , \ \ {\cal{R}} = 1.67 \ .
\end{equation}
Because of these results, we can therefore conclude that including
a negative $\Lambda$ leads to a model still in agreement with the data so
that this approach to halting eternal acceleration is a viable one from an
observational point of view. \footnote{see \cite{Cardone2008} for a further discussion about the observational results and implications}

Another interesting tools to study the viability of a dark energy model is the point of view of structure formation. This kind of analysis can break beetwen models with similar prediction from the cosmic expansion history, in this sense the growth of the large scale structure in the universe provide  and important companion test. Following this line in \cite{Leyva2009} the authors showed that the quinstant model makes reasonable predictions for the formation of linear large scale structure of the Universe but it fails in the non linear regime because of the density contrast at virialisation increase with the value of virialisation redshif. 

Concerning the predictions of the cluster abundances, the quinstant model is capable of reproducing the results of the other models in a satisfactory way backwards in time up to redshifts a bit larger than $z=1$ for the three range of mass values \footnote{the same behavior is obtained for other mass range \cite{Leyva2009}}. Then, it shows abrupt peaks of structure formation, in a serious departure of the hierarchical model for large scale structure. This seems to be caused by the unusual equation of state of quinstant dark energy, which behaves as stiff matter for redshifts a bit larger than one. This would result in enhanced accretion of the forming structures, both because of gravitational and viscous forces.

\section{Exponential Quinstant: Phase space analysis}

In this section we will investigate, from the dynamical systems
viewpoint the quinstant dark energy model with expoenential
potential $V(\phi)=V_0 e^{\lambda\phi}.$ We do not consider
radiation fluid here but a background of a perfect fluid with
equation of state $w=\gamma-1.$ The cosmological equations reads:

\begin{eqnarray}
&
H^2-\frac{1}{6}\dot\phi^2-\frac{1}{3}V(\phi)-\frac{1}{3}\rho_{\rm
M}-\frac{\Lambda}{3}=-\frac{k}{a^2},\;
k=-1,0,1,\\
& \dot H=-H^2-\frac{1}{3}\dot\phi^2+\frac{1}{3}V(\phi)+\frac{\Lambda}{3}-\frac{1}{6}\left(3\gamma-2\right)\rho_{\rm M},\\
& \dot\rho_{\rm M}=-3 \gamma H \rho_{\rm M},\\
& \ddot\phi+3 H \dot\phi+\frac{d V(\phi)}{d\phi}=0.
\end{eqnarray}

Our purpose is to re-express the former equations as an autonomous
dynamical system.

\subsection{Flat FRW case}\label{Flat}

\subsubsection{Normalization, state space, and dynamical system.}\label{Normalization}

In order to get a first order autonomous system of ordinary
differential equations (ODEs) is is convenient introduce
normalized variables an a new convenient (monotonic) time
variable. If the phase space is compact the flow of the system
admits both past and future attractors. Let us introduce the
normalization factor $D=\sqrt{H^2-\Lambda/3}>0,$ the new time
variable $d\tau=D dt,$ and the phase space variables:
\begin{equation}
x=\frac{\dot\phi}{\sqrt{6} D},\; y=\frac{\sqrt{V(\phi)}}{\sqrt{3}
D},\; {\cal H}=\frac{H}{D}.\label{Vars}
\end{equation}
The former variables lies in the compact phase space
$$\left\{(x,y,z): x^2+y^2\leq 1,\; y\geq 0,\; -1\leq {\cal H}
\leq 1\right\}.$$

The variables (\ref{Vars}) satisfy the ODEs (the prime denoting
derivative with respect to $\tau$):
\begin{eqnarray}
x'&=&-\frac{3}{2} x {{\cal H}} \left((\gamma -2)
x^2+\left(y^2-1\right) \gamma
   +2\right)\nonumber\\ && -\sqrt{\frac{3}{2}} \lambda y^2,\label{Eqx}\\
   y'&=&\frac{3}{2} y \left(\frac{\sqrt{6} \lambda x}{3}-{{\cal H}} \left((\gamma -2) x^2+\left(y^2-1\right) \gamma \right)\right),\label{Eqy}\\
   {{\cal H}}'&=&-\frac{3}{2} \left({{\cal H}}^2-1\right) \left((\gamma -2) x^2+\left(y^2-1\right) \gamma \right)\label{Eqz}
\end{eqnarray}

For convenience, let us express some cosmological quantities in
terms of the variables (\ref{Vars}). The deceleration parameter is
explicitly
\begin{equation} q\equiv-\ddot a a/\dot a^2=-1+\frac{3}{2}\left[\frac{x^2 \left(2-\gamma\right)+ \left(1-y^2\right) \gamma }{{\cal H}^2}\right];\end{equation} the fractional energy density of the scalar field is
\begin{equation}
\Omega_\phi=\frac{x^2+y^2}{{{\cal H}}^2};
\end{equation}
and the 'effective' EoS parameter is given by
\begin{equation}
\omega_{eff}\equiv\frac{P_{tot}}{\rho_{tot}}\equiv\frac{\frac{1}{2}
\dot\phi^2-V(\phi )+(\gamma -1) \rho_{\rm M} -\Lambda
}{\frac{1}{2} \dot\phi ^2+V(\phi )+\rho_{\rm M} +\Lambda
}=-1+\frac{\left(2-\gamma\right) x^2+\left(1-y^2\right) \gamma
}{{\cal H}^2}.
\end{equation}

\subsubsection{Form invariance under coordinate transformations.}

The system (\ref{Eqx}-\ref{Eqz}) is form invariant under the
coordinate transformation and time reversal

\begin{equation}\left(\tau,\,x,\,y,\,{\cal H}\right)\rightarrow
\left(-\tau,\,-x,\,y,\,-{\cal
H}\right).\label{discrete3}\end{equation}

Thus, it is sufficient to discuss the behaviour in one part of the
phase space, the dynamics in the other part being obtained via the
transformation (\ref{discrete3}). Observe also, that equations
(\ref{Eqx}-\ref{Eqz}) are form invariant under the coordinate
transformation $y\rightarrow -y.$ Then, (\ref{Eqx}-\ref{Eqz}) is
form invariant under its composition with (\ref{discrete3}).

From equation (\ref{Eqz}) follows that ${\cal H}=\pm 1$ are
invariant sets of the flow. From equation (\ref{Eqy}) follows that
the sign of $y$ is invariant.

\subsubsection{Monotonic functions.}

Let be defined \begin{equation} Z(x,y,{\cal H})=\left(\frac{{\cal
H}+1}{{\cal H}-1}\right)^2\label{Z}\end{equation} in the invariant
set $$S=\left\{(x,y,{\cal H}): x^2+y^2< 1,\; y> 0,\; -1< {\cal H}
< 1\right\}.$$ Then, $Z$ is monotonic decreasing in $S$ since
$$Z'\equiv\nabla Z\cdot f=-6\,Z\,\left( x^2\,\left(2 - \gamma
\right)  + \left( 1 - y^2 \right) \,\gamma \right)<0$$ in $S.$ The
existence of this monotonic allows to state that there can be no
periodic orbits or recurrent orbits in the interior of the phase
space. Furthermore, it is possible to obtain global results. The
range of $Z$ is the semi-interval $(0,+\infty),$  and
$Z\rightarrow 0$ as ${\cal H}\rightarrow -1$ (since ${\cal H}$ is
bounded) and $Z\rightarrow +\infty$ as ${\cal H}\rightarrow 1.$ By
applying the Monotonicity Principle (theorem 4.12 \cite{REZA}) we
find that, for all $p\in S,$ the past asymptotic attractor of $p$
(the $\alpha$-limit) belongs to ${\cal H}=1$ and the future
asymptotic attractor of $p$ (the $\omega$-limit) belongs to ${\cal
H}=-1.$

\subsubsection{Local analysis of critical points.}\label{local}

The system (\ref{Eqx}-\ref{Eqy}) admits ten critical points with
the labels $P_i^{\pm}$ with $i=1\ldots 5.$ In table \ref{tab1} we
offer some partial information about the location, conditions for
existence and some additional properties of them. All the critical
points satisfy ${\cal H}=\pm 1.$ In other words, they are
solutions with $H=\pm D$ (i.e. with $H\rightarrow\pm \infty$). If
$\text{sign}({\cal H})=-1$ the associated solutions ends in a
collapse (since $H<0$), whereas, if $\text{sign}({\cal H})=1$ we
have ever expanding cosmological solutions. The expected
cosmological behavior of our model is that the attractor solutions
represent collapsing solutions due the negative value of the
cosmological constant.

\begin{table*}[ht]\caption[crit]{Location and existence conditions of the critical points of the system
(\ref{Eqx}-\ref{Eqz})}
\begin{center}
\begin{tabular}{@{\hspace{4pt}}c@{\hspace{14pt}}c@{\hspace{14pt}}c@{\hspace{2pt}}}
\hline
\hline\\[-0.3cm]
Label & Coordinates: $(x,y,{\cal H})$& Existence\\[0.1cm]
\hline\\[-0.2cm]
$P_1^\pm$& $(-1,0,\pm 1)$& All $\lambda$  \\\hline\\
$P_2^\pm$& $(0,0,\pm 1)$& All $\lambda$  \\\hline\\
$P_3^\pm$ & $(1,0,\pm 1)$& All $\lambda$   \\\hline\\
$P_4^\pm$&
$\left(\mp\frac{\lambda}{\sqrt{6}},\sqrt{1-\frac{\lambda^2}{6}},\pm 1\right)$& $-\sqrt{6}< \lambda<\sqrt{6}$ \\\hline\\
$P_5^\pm$&
$\left(\mp\sqrt{\frac{3}{2}}\frac{\gamma}{\lambda},\sqrt{\frac{3}{2}}
\sqrt{\frac{(2-\gamma)\gamma}{\lambda^2}},\pm 1\right)$ &
$\begin{array}{c}
\gamma=0,\;\lambda\neq 0\\[0.2cm]
0<\gamma\leq 2,\; |\lambda|  \geq \sqrt{3\gamma}\\[0.2cm]
\end{array} $
\\[0.4cm]
\hline \hline
\end{tabular}\label{tab1}
\end{center}
\end{table*}

\begin{table*}[ht]\caption[crit2]{Eigenvalues, and dynamical character of the fixed points of (\ref{Eqx}-\ref{Eqz}).
We use the notation
$\Delta=(2-\gamma)(24\gamma^2+\lambda^2(2-9\gamma)).$}
\begin{center}
\begin{tabular}{@{\hspace{2pt}}c@{\hspace{10pt}}c@{\hspace{14pt}}c@{\hspace{2pt}}}
\hline
\hline\\[-0.3cm]
Label & Eigenvalues& Dynamical character\\[0.1cm]
\hline\\[-0.2cm]
$P_1^-$& $- 6,- 3-\sqrt{\frac{3}{2}}\lambda,- 3(2-\gamma)$& $\begin{array}{c} \text{nonhyperbolic if}\; \gamma=2\; \text{or}\;  \lambda=-\sqrt{6};\\[0.1cm]
\text{stable (node) if}\; \lambda>-\sqrt{6}\; \text{and}\; \gamma\neq 2;\\[0.1cm]
\text{saddle, otherwise.}\\[0.1cm]\end{array}$\\\hline
$P_2^-$& $\frac{3}{2}(2-\gamma),- 3\gamma,- \frac{3\gamma}{2}$&  $\begin{array}{c} \text{nonhyperbolic if}\; \gamma=0\; \text{or}\;  \gamma=2;\\[0.1cm]
\text{saddle, otherwise.}\\[0.1cm]\end{array}$ \\\hline
$P_3^-$& $- 6,- 3+\sqrt{\frac{3}{2}}\lambda,- 3(2-\gamma)$& $\begin{array}{c} \text{nonhyperbolic if}\; \gamma=2\; \text{or}\;  \lambda=\sqrt{6};\\[0.1cm]
\text{stable (node) if}\; \lambda<\sqrt{6}\; \text{and}\; \gamma\neq 2;\\[0.1cm]
\text{saddle, otherwise.}\\[0.1cm]\end{array}$ \\\hline
$P_4^-$& $-\lambda^2,\frac{1}{2}(6-\lambda^2),-\lambda^2+3\gamma$& $\begin{array}{c} \text{nonhyperbolic if}\; \lambda=0\; \text{or}\;  \lambda^2=3\gamma;\\[0.1cm]
\text{saddle, otherwise.}\\[0.1cm]\end{array}$ \\\hline
$P_5^-$& $-3\gamma,\frac{3}{4} \left(2-\gamma
\pm\frac{1}{\lambda}\sqrt{\Delta}\right)$& $\begin{array}{c} \text{nonhyperbolic if}\; \gamma=0\; \text{or}\;  \lambda^2=3\gamma;\\[0.1cm]
\text{saddle, otherwise.}\\[0.1cm]\end{array}$
\\[0.3cm]
\hline \hline
\end{tabular}\label{tab2}
\end{center}
\end{table*}

\begin{table*}[ht]\caption[crit]{Some properties of the critical points of the system
(\ref{Eqx}-\ref{Eqz})}
\begin{center}
\begin{tabular}{@{\hspace{4pt}}c@{\hspace{14pt}}c@{\hspace{18pt}}c@{\hspace{18pt}}c@{\hspace{2pt}}}
\hline
\hline\\[-0.3cm]
Label & Deceleration $q$& $\Omega_\phi$& $\omega_{\text{eff}}$\\[0.1cm]
\hline\\[-0.2cm]
$P_1^\pm$ & $2$& $1$ & $1$\\\hline\\
$P_2^\pm$ & $-1+\frac{3\gamma}{2}$& $0$ & $-1+\gamma$\\\hline\\
$P_3^\pm$ & $2$&
$1$ & $1$ \\\hline\\
$P_4^\pm$& $-1+\frac{\lambda^2}{2}$ &$1$&$-1+\frac{\lambda^2}{3}$ \\\hline\\
$P_5^\pm$ & $-1+\frac{3\gamma}{2}$ & $\frac{3\gamma}{\lambda^2}$&
$-1+\gamma$
\\[0.4cm]
\hline \hline
\end{tabular}\label{tab1b}
\end{center}
\end{table*}

Now, let us make some comments about the cosmological solutions
associated to these critical points.

The critical points $P_1^\pm$ and $P_3^\pm$ represent stiff-matter
solutions which are associated with massless scalar field
cosmologies (the kinetic energy density of the scalar field
dominated against the potential energy density). In the former
case the scalar field is a monotonic decreasing function of $t$
(since its time-derivative is negative). In the last case the
scalar field is an increasing function of $t$ since its
time-derivative is positive. These solutions are always
decelerated. The critical points $P_2^\pm$ represent a flat FRW
solution fuelled by perfect fluid. They represent accelerating
solutions for $\gamma<\frac{2}{3}.$ The critical points $P_4^\pm$
represent solutions dominated by the scalar field
($\Omega_\phi=1$, and $H\rightarrow\pm\infty$) which are
accelerating if $\lambda^2<2.$ Our models does not devoid of
scaling phases: the critical points $P_5^\pm$ are such that
neither the scalar field nor the perfect fluid dominates the
evolution. There $\Omega_m/\Omega_\phi=\text{const.},$ and
$\gamma_\phi=\gamma.$

Before proceed to make some numerical experiments let us discuss
some aspects concerning the symmetry (\ref{discrete3}). Observe
that the critical points $P_3^\mp,$  $P_2^\mp,$ $P_4^\mp$ and
$P_5^\mp$ are related by the transformation (\ref{discrete3}) with
$P_1^\pm,$ $P_2^\pm,$ $P_4^\pm$ and $P_5^\pm$ respectively. In
order to analyze the local stability of $P_1^+, P_2^+, P_3^+,
P_4^+, P_5^+$ it is sufficient analyse the local stability of
$P_3^-, P_2^-, P_1^-, P_4^-, P_5^-$ respectively, and then, infer
the stability of the points in the ``positive'' branch by using
(\ref{discrete3}). In table \ref{tab2} we offer partial
information about the dynamical character of the critical points
corresponding to the ``negative'' branch.

The critical point $P_1^-$ is nonhyperbolic provided $\gamma=2$ or
$\lambda=-\sqrt{6}.$ it is a stable node (future attractor)
provided $\lambda>-\sqrt{6}$ and $\gamma\neq 2.$ It is a saddle
otherwise with a 2D stable manifold and a 1D unstable manifold
tangent to the $y$-axis. $P_2^-$ is nonhyperbolic provided
$\gamma=0$ or $\gamma=2.$ It is a saddle point otherwise with a 2D
stable manifold and a 1D unstable manifold tangent to the
$x$-axis. The critical point $P_3^-$ is nonhyperbolic provided
$\gamma=2$ or $\lambda=\sqrt{6}.$ it is a stable node (future
attractor) provided $\lambda<\sqrt{6}$ and $\gamma\neq 2.$ It is a
saddle otherwise with a 2D stable manifold and a 1D unstable
manifold tangent to the $y$-axis. $P_4^-$ is nonhyperbolic if
$\lambda^2\in\{0,3\gamma, 6\}.$ Saddle otherwise, with a 2D stable
manifold provided $\lambda^2>3\gamma$ or 1D if
$\lambda^2<3\gamma.$ The critical point $P_5^-$ is nonhyperbolic
if $\gamma=0$ or $\lambda^2=3\gamma.$ It is a saddle point,
otherwise, with a 2D unstable manifold provided $0<\gamma<2$ and
$\lambda^2>3\gamma.$ $P_4^-$ ($P_5^-$) is the past attractor in
the invariant set ${\cal H}=-1$ provided $0<\gamma<2,\,
\lambda^2<3\gamma$ ($0<\gamma<2,\, \lambda^2>3\gamma$).

In table \ref{table9}, where we present a summary of attractors
(both past and future) for the quinstant model with $k=0.$

\begin{table*}[ht!]\caption{Summary of attractors for for the quinstant model with $k=0$ (system (\ref{Eqx}-\ref{Eqz})).}\label{table9}
\begin{tabular}{@{\hspace{4pt}}c|@{\hspace{10pt}}c@{\hspace{10pt}}c@{\hspace{2pt}}}
\hline
\hline\\[-0.3cm]
Restrictions &  Past attractor    & Future attractor \\[0.1cm]
\hline\\[-0.3cm]
${\cal H}=-1$ & $\begin{array}{cc}
P_4^- & \text{if}\;0<\gamma<2,\, \lambda^2<3\gamma\\[0.2cm]
P_5^- & \;0<\gamma<2,\, \lambda^2>3\gamma\\[0.2cm]
\end{array}$   & $\begin{array}{cc}
P_3^-& \text{if}\;\lambda<\sqrt{6},\,\gamma\neq 2\\[0.2cm]
P_1^- & \text{if}\;\lambda>-\sqrt{6},\,\gamma\neq 2\\[0.2cm]
\end{array}$  \\[0.2cm]\hline\\[0.1cm]
$-1<{\cal H}<1$ & $\begin{array}{cc}
P_3^+ & \text{if}\;\lambda>-\sqrt{6},\,\gamma\neq 2\\[0.2cm]
P_1^+& \text{if}\;\lambda<\sqrt{6},\,\gamma\neq 2\\[0.2cm]
\end{array}$  & as above \\[0.2cm]\hline\\[0.1cm]
${\cal H}=1$ & as above  & $\begin{array}{cc}
P_4^+ & \text{if}\;0<\gamma<2,\, \lambda^2<3\gamma\\[0.2cm]
P_5^+ & \;0<\gamma<2,\, \lambda^2>3\gamma\\[0.2cm]
\end{array}$ \\[0.4cm]
\hline \hline
\end{tabular}
\end{table*}

\begin{figure}[!]
\begin{center}
\hspace{0.4cm} \put(140,195){${P_1^+}$} \put(140,75){${P_1^-}$}
\put(105,170){${P_2^+}$} \put(105,45){${P_2^-}$}
\put(70,135){${P_3^+}$} \put(70,15){${P_3^-}$}
\put(125,143){${P_4^+}$} \put(155,45){${P_4^-}$}
\put(30,45){${x}$} \put(195,80){${y}$} \put(105,195){${{\cal H}}$}
\mbox{\epsfig{figure=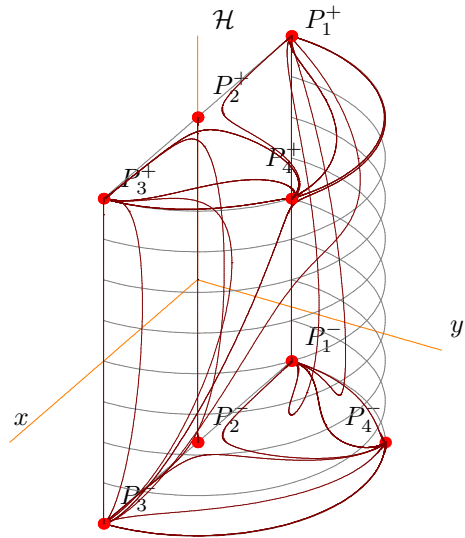,width=7cm,angle=0}}
\caption{(Color online) {\it Some orbits of (\ref{Eqx}-\ref{Eqy})
in the phase space for the values $\lambda=-\sqrt{\frac{3}{2}}$
and $\gamma=1.$ The critical points $P_1^-$ (resp. $P_1^+$) and
$P_3^-$ (resp. $P_3^+$) are the future (resp. past) asymptotic
attractors, $P_3^-$ (resp. $P_1^+$) having a stronger attracting
(resp. unstable) manifold tangent to the $y$-axis. $P_4^+,$ acts
as a local attractor in the invariant set ${\cal H}=1$ and
$P_4^-,$ acts as the local source for the invariant set ${\cal
H}=-1$ (they are, however, saddle points for the 3D
dynamics).}}\label{Phase2}
\end{center}
\end{figure}

In figure \ref{Phase2} we show some orbits in the phase space for
the values $\lambda=-\sqrt{\frac{3}{2}}$ and $\gamma=1.$ For this
choice $\lambda^2<3\gamma$ and $-\sqrt{6}<\lambda<\sqrt{6}.$ Thus
the critical points $P_5^\pm$ do not exist.  By the linear
analysis (see table \ref{tab2}) we find that the critical points
$P_1^-$ and $P_3^-$ have a 3-dimensional stable manifold, $P_3^-$
having a stronger attracting manifold tangent to the $y$-axis (see
figure \ref{Phase2}), i.e., two global future attractors might
coexist (bistability). The critical points $P_1^+$ and $P_3^+$ are
local sources in the invariant set ${\cal H}=1$. Numerical
inspection suggest and analytical results confirm that they are
also global sources, $P_1^+$ having a stronger unstable direction
tangent to $y$-axis. The critical points $P_2^\pm$ acts locally as
saddles. For $P_2^-$ (resp. $P_2^+$) the stable (resp. unstable)
manifold is 2-dimensional and tangent to the y-${\cal H}$ plane.
There are orbits (corresponding to exact cosmological solutions)
connecting $P_{1,2,3}^+$ with $P_{1,2,3}^-$ (recollapse occurs).
The critical point $P_4^+,$ with coordinates $(1/2,\sqrt{3}/2,
1),$ have eigenvalues $-2.25, -1.5, 1.5$ acting as a local
attractor in the invariant set ${\cal H}=1$ and $P_4^-,$ with
coordinates $(-1/2,\sqrt{3}/2, 1),$ and eigenvalues $2.25, 1.5,
-1.5$ is the local source for the invariant set ${\cal H}=-1$.
They are saddle points for the 3D dynamics (see figure
\ref{Phase2}).

\subsubsection{Bifurcations.}

The critical points $\left(P_4^+,\, P_4^-\right)$ reduce to
$\left(P_1^+,\, P_3^-\right)$ as $\lambda\rightarrow
(\sqrt{6})^-.$ The critical points $\left(P_4^+,\, P_4^-\right)$
reduce to $\left(P_3^+,\, P_1^-\right)$ as $\lambda\rightarrow
(-\sqrt{6})^+.$ The critical points $P_5^\pm$ reduce to $P_2^\pm$
as $\gamma\rightarrow 0^+.$ On the other hand, $P_5^\pm$ reduce to
$P_4^\pm$ as $\lambda\rightarrow (\sqrt{3\gamma})^+$ or
$\lambda\rightarrow (-\sqrt{3\gamma})^-.$ For these values of the
parameters a bifurcation arises.

\subsubsection{Typical behavior}

Once the attractors have been identified one can give a
quantitative description of the physical behavior of a typical
flat quinstant cosmology.

For example, for ever expanding cosmologies with $H>0,
H\rightarrow +\infty, {\cal H}=1$, i.e., the standard expanding
cosmology near the big-bang, a typical model behaves like a
massless scalar field (kinetic dominated energy density)
represented by $P_3^+$ or $P_1^+$ provided $\lambda>-\sqrt{6},\,
\gamma\neq 2$ or $\lambda<\sqrt{6},\, \gamma\neq 2,$ respectively.
This types of solutions might coexist in the same phase space. The
late time dynamics in ${\cal H}=1$ is given by either a scalar
field dominated solution ($\Omega_\phi\rightarrow 1$) represented
by $P_4^+$ or by a scaling solution ($\Omega_m/\Omega_\phi=O(1)$)
represented by $P_5^+$ provided $\lambda^2<3\gamma$ or
$\lambda^2>3\gamma,$ respectively. For finite values of $H$, i.e.,
$-1<{\cal H}<1,$ the early time dynamics is the same as in the
previous case but there are subtle differences with respect the
late time dynamics. In fact, in the invariant set $-1<{\cal H}<1$
the future attractors are $P_3^-$ or $P_1^-$ depending if
$\lambda<\sqrt{6},\, \gamma\neq 2$ or $\lambda>-\sqrt{6},\,
\gamma\neq 2.$ If $|\lambda|<\sqrt{6}$ the system is bistable.
Such solutions represent contracting stiff-fluid cosmologies. This
means that a typical quinstant cosmologies allows the collapse of
matter when the time evolves. For contracting cosmologies with
($H<0, H\rightarrow\-\infty, {\cal H}=-1$), i.e., the standard
contracting model near the initial singularity, the late time
dynamics is the same as int the previously described case, i.e.,
the collapse. However, there are subtle differences concerning the
early time dynamics. The late time dynamics in ${\cal H}=-1$ is
given by either a scalar field dominated solution
($\Omega_\phi\rightarrow 1$) represented by $P_4^-$ or by a
scaling solution ($\Omega_m/\Omega_\phi=O(1)$) represented by
$P_5^-$ provided $\lambda^2<3\gamma$ or $\lambda^2>3\gamma.$


\subsection{Quinstant cosmology with negative
curvature}\label{quinstantneg}

\subsubsection{Normalization, state space, and dynamical system.}

Let us consider the same normalization as in section
\ref{Normalization}, i.e, the normalization factor
$D=\sqrt{H^2-\Lambda/3}>0$ and the time variable $d\tau=D dt.$ We
will consider the variables (\ref{Vars}) augmented by the new
variable $z=\frac{1}{a D}.$ These variables lies in the compact
phase space
$$\left\{(x,y,z,{\cal H}): x^2+y^2+z^2\leq 1,\; y\geq 0,\; z\geq 0,\; -1\leq {\cal H}
\leq 1\right\}.$$
The variables $x,\,y,\,z,$ and ${\cal H}$ satisfy the ASODE (the
prime denoting derivative with respect to $\tau$):

\begin{eqnarray}
x'&=&-\frac{3}{2} x {{\cal H}} \left((\gamma -2)
x^2+\left(y^2-1\right) \gamma
   +z^2\left(\gamma-\frac{2}{3}\right)+2\right)\nonumber\\ && -\sqrt{\frac{3}{2}} \lambda y^2,\label{negEqx}\\
   y'&=&\frac{3}{2} y \left(\frac{\sqrt{6} \lambda x}{3}-{{\cal H}} \left((\gamma -2) x^2+\left(y^2-1\right) \gamma + z^2\left(\gamma-\frac{2}{3}\right)\right)\right),\label{negEqy}\\
   z'&=&\frac{3}{2}z {\cal H}\left((\gamma -2) x^2+\left(y^2-1\right) \gamma + z^2\left(\gamma-\frac{2}{3}\right)-\frac{2}{3}\right)\label{negEqz}\\
   {{\cal H}}'&=&-\frac{3}{2} \left({{\cal H}}^2-1\right) \left((\gamma -2) x^2+\left(y^2-1\right) \gamma + z^2\left(\gamma-\frac{2}{3}\right)\right)\label{negEqH}
\end{eqnarray}

As before, we will re-express the cosmological magnitudes of
interest in terms of the normalized variables.

The deceleration parameter is explicitly
\begin{equation} q\equiv-\ddot a a/\dot a^2=-1+\frac{3}{2}\left[\frac{x^2 \left(2-\gamma\right)+ \left(1-y^2\right) \gamma +z^2\left(\frac{2}{3}-\gamma\right)}{{\cal H}^2}\right];\end{equation}
the fractional energy density of the scalar field and curvature
are given respectively by

\begin{equation}
\Omega_\phi=\frac{x^2+y^2}{{{\cal H}}^2};
\Omega_k=\frac{z^2}{{\cal H}^2}
\end{equation}
and the 'effective' EoS parameter is given by

\begin{equation}
\omega_{eff}\equiv\frac{P_{tot}}{\rho_{tot}}\equiv\frac{\frac{1}{2}
\dot\phi^2-V(\phi )+(\gamma -1) \rho_{\rm M} -\Lambda
}{\frac{1}{2} \dot\phi ^2+V(\phi )+\rho_{\rm M} +\Lambda
}=-1+\frac{\left(2-\gamma\right) x^2+\left(1-y^2\right) \gamma
-\gamma z^2}{{\cal H}^2-z^2}.
\end{equation}

\subsubsection{Form invariance under coordinate transformations.}

The system (\ref{negEqx}-\ref{negEqH}) is form invariant under the
coordinate transformation and time reversal

\begin{equation}\left(\tau,\,x,\,y,\,z,\,{\cal H}\right)\rightarrow
\left(-\tau,\,-x,\,y,\,z,\,-{\cal
H}\right).\label{discrete4}\end{equation}

Thus, it is sufficient to discuss the behaviour in one part of the
phase space, the dynamics in the other part being obtained via the
transformation (\ref{discrete4}). Observe that equations
(\ref{negEqx}-\ref{negEqH}) are form invariant under the
coordinate transformation $y\rightarrow -y$ and $z\rightarrow -z.$
Then, (\ref{negEqx}-\ref{negEqH}) is form invariant under they
composition with (\ref{discrete4}).

There are four obvious invariant sets under the flow of
(\ref{negEqx}-\ref{negEqH}), they are $y=0,$ $z=0,$ and ${\cal
H}=\pm 1.$ They combination defines other invariant sets. The
dynamics restricted to the invariant set $z=0$ is the same
described in section \ref{Flat}. It is of course of interest the
analysis of the behavior of the 4D orbits near the invariant set
$z=0$ (or the other enumerated above). We will discuss this in
next sections.

\subsubsection{Monotonic functions.}

Let be defined \begin{equation} Z_1=\left(\frac{{\cal H}+1}{{\cal
H}-1}\right)^2\label{Z1}\end{equation} in the invariant set
$$\left\{(x,y,z, {\cal H}): x^2+y^2+z^2< 1,\; y> 0,\; z> 0,\; -1<{\cal H}<
1\right\}.$$ Then, $Z$ is monotonic decreasing in $S$ since
$$Z'\equiv\nabla Z\cdot f=-Z\,\left(4 z^2+ 6 x^2\,\left(2 - \gamma
\right)  + 6\left( 1 - y^2 -z^2\right) \,\gamma \right)<0$$ in
$S.$ The range of $Z$ is the semi-interval $(0,+\infty),$  and
$Z\rightarrow 0$ as ${\cal H}\rightarrow -1$ (since ${\cal H}$ is
bounded) and $Z\rightarrow +\infty$ as ${\cal H}\rightarrow 1.$ By
applying the Monotonicity Principle (theorem 4.12 \cite{REZA}) we
find that, for all $p\in S,$ the past asymptotic attractor of $p$
(the $\alpha$-limit) belongs to ${\cal H}=1$ and the future
asymptotic attractor of $p$ (the $\omega$-limit) belongs to ${\cal
H}=-1.$

Let be defined in the same invariant set the function

\begin{equation}
Z_2=\frac{z^4}{\left(1-x^2-y^2-z^2\right)^2},\;
Z_2'=-2\left(2-3\gamma\right){\cal H}Z_2\label{Z2}
\end{equation} This function is monotonic in the regions ${\cal H} < 0$ and ${\cal H}>0$ for $\gamma\neq \frac{2}{3}.$

The existence of monotonic functions allows to state that there
can be no periodic orbits or recurrent orbits in the interior of
the phase space. Futhermore, it is possible to obtain global
results. From the expression $Z_2$ we can immediately see that
asymptotically $z\rightarrow 0$ or $x^2+y^2+z^2\rightarrow 1.$

\subsubsection{Local analysis of critical points.}

The system (\ref{negEqx}-\ref{negEqH}) admits fourteen critical
points. We will denote the critical points of the system
(\ref{negEqx}-\ref{negEqH}) located at the invariant set $z=0$ in
the same way as in the table \ref{tab1} of section \ref{local}. We
submit the reader to this table for the conditions for their
existence. In table \ref{negtab2} are summarized the stability
properties of the critical points.

Observe that the critical points $P_6^\mp$ and $P_7^\pm$ are
related through (\ref{discrete4}) with the critical points
$P_6^\pm$ and $P_7^\pm$ respectively.

In order to analyze the local stability of $P_1^+, P_2^+, P_3^+,
P_4^+, P_5^+, P_6^+, P_7^+$ it is sufficient analyze the local
stability of $P_3^-, P_2^-, P_1^-, P_4^-, P_5^-,P_6^-, P_7^-$
respectively, and then, infer the stability of the points in the
``positive'' branch by using (\ref{discrete4}). In table
\ref{negtab2} we offer a detailed analysis of the dynamical
character of the critical points corresponding to the ``negative''
branch.

The critical point ${P}_1^-$ is nonhyperbolic if $\gamma=2$ or
$\lambda=-\sqrt{6}.$ It is a local sink provided
$\lambda>-\sqrt{6},\gamma\neq 0.$ If $\lambda<-\sqrt{6}$ then
there exists a 1D unstable manifold tangent to the $\hat{y}$-axis,
and a 3D stable manifold. The critical point ${P}_2^-$ is
nonhyperbolic if $\gamma=0,$ $\gamma=\frac{2}{3}$ or $\gamma=2.$
There exist always at least a 1D unstable manifold tangent to the
${x}$ axis.  The unstable manifold is 2D provided
$\gamma<\frac{2}{3}.$ In this case there exists a 2D stable
manifold tangent to the ${y}$-${\cal H}$ plane. The critical point
${P}_3^-$ is nonhyperbolic if $\gamma=2$ or $\lambda=\sqrt{6}.$ It
is a local sink provided $\lambda<\sqrt{6},\gamma\neq 0.$ If
$\lambda>\sqrt{6}$ then there exists a 1D unstable manifold
tangent to the ${y}$-axis, and a 3D stable manifold. Observe that
${P}_1^-$ and ${P}_3^-$ coexist provided $\lambda^2<6.$ In this
case both are the future attractors of the system, they attract
solutions in its basin of attraction. The critical point
${P}_4^-,$ is hyperbolic if $\lambda\in\{0, \pm 3\gamma, \pm 2\}$.
If not, it is always a saddle point with an unstable manifold at
least 1D. The stable manifold is 3D provided $0<
\gamma\leq\frac{2}{3}, \, 2<\lambda^2<6$ or
$\frac{2}{3}<\gamma<2,\, 3\gamma<\lambda^2<6.$ It is 2D provided
$0< \gamma<\frac{2}{3}, \, 3\gamma<\lambda^2<2$ or
$\frac{2}{3}<\gamma< 2,\, 2<\lambda^2<3\gamma$ or 1D provided
$0<\gamma\leq \frac{2}{3},\, 0<\lambda^2<3\gamma$ or
$\frac{2}{3}<\gamma<2,\, 0<\lambda^2<2.$ The critical point
$P_5^-$ is nonhyperbolic if  $\gamma=0$ or $\lambda^2=3\gamma$ or
$\gamma=\frac{2}{3}.$ If not, it have always two conjugate complex
eigenvalues with positive real parts (there exists at least a 2D
unstable manifold). The stable manifold is 1D provided
$0<\gamma\leq \frac{2}{9},\, \lambda^2>3\gamma$ or
$\frac{2}{9}<\gamma<\frac{2}{3},\, \lambda^2<\frac{24
\gamma^2}{-2+9\gamma}$ or $2D$ if $\frac{2}{3}<\gamma<2,\,
\lambda^2<\frac{24 \gamma^2}{-2+9\gamma}.$ Thus, $P_5^-$ is always
a saddle point.

In the negative curvature case there are two new classes of
critical points: Mine's solutions and curvature-scaling solutions,
denoted by $P_6^\pm$ and $P_7^\pm$ respectively. In table
\ref{negtab1} it is displayed the location, existence and some
properties of them.

\begin{table*}[ht]\caption[crit]{Location and existence conditions of the critical points
$P_6^\pm$ and $P_7^\pm$ of the system
(\ref{negEqx}-\ref{negEqH}).}
\begin{center}
\begin{tabular}{@{\hspace{4pt}}c@{\hspace{14pt}}c@{\hspace{14pt}}c@{\hspace{2pt}}}
\hline
\hline\\[-0.3cm]
Label & Coordinates: $(x,y,z,{\cal H})$& Existence\\[0.1cm]
\hline\\[-0.2cm]
$P_6^\pm$& $(0,0,1,\pm 1)$& All $\lambda$ \\[0.1cm]\hline
$P_7^\pm$& $(\mp
\frac{1}{\lambda}\sqrt{\frac{2}{3}},\frac{2}{\sqrt{3}\lambda},\frac{1}{\lambda}\sqrt{-2+\lambda^2},\pm
1)$& $\begin{array}{c} \frac{2}{3}<\gamma\leq 2\\[0.1cm]
\text{and}\; |\lambda|\geq \sqrt{2}\\[0.1cm]\end{array}$
\\[0.4cm]
\hline \hline
\end{tabular}\label{negtab1}
\end{center}
\end{table*}

\begin{table*}[ht]\caption[crit]{Some properties of of the critical points
$P_6^\pm$ and $P_7^\pm$ of the system
(\ref{negEqx}-\ref{negEqH}).}
\begin{center}
\begin{tabular}{@{\hspace{4pt}}c@{\hspace{14pt}}c@{\hspace{18pt}}c@{\hspace{18pt}}c@{\hspace{2pt}}}
\hline
\hline\\[-0.3cm]
Label & Deceleration $q$& $\Omega_\phi$& $\omega_{\text{eff}}$\\[0.1cm]
\hline\\[-0.2cm]
$P_6^\pm$& $0$ & $1$ & $0$\\[0.1cm]\hline
$P_7^\pm$ & $\frac{2}{\lambda^2}$ & $1-\frac{2}{\lambda^2}$ &
$-\frac{1}{3}$
\\[0.4cm]
\hline \hline
\end{tabular}\label{negtab1b}
\end{center}
\end{table*}

\begin{table*}[ht]\caption[crit2]{Eigenvalues, and dynamical character of the fixed points of the system (\ref{negEqx}-\ref{negEqH}).
We use the notation
$\Delta=(2-\gamma)(24\gamma^2+\lambda^2(2-9\gamma)).$}
\begin{center}
\begin{tabular}{@{\hspace{2pt}}c@{\hspace{10pt}}c@{\hspace{14pt}}c@{\hspace{2pt}}}
\hline
\hline\\[-0.3cm]
Label & Eigenvalues& Dynamical character\\[0.1cm]
\hline\\[-0.2cm]
$P_1^-$& $- 6,- 3-\sqrt{\frac{3}{2}}\lambda,- 3(2-\gamma), -2$& $\begin{array}{c} \text{nonhyperbolic if}\; \gamma=2\; \text{or}\;  \lambda=-\sqrt{6};\\[0.1cm]
\text{stable (node) if}\; \lambda>-\sqrt{6}\; \text{and}\; \gamma\neq 2;\\[0.1cm]
\text{saddle, otherwise.}\\[0.1cm]\end{array}$\\\hline
$P_2^-$& $\frac{3}{2}(2-\gamma),- 3\gamma,- \frac{3\gamma}{2},1-
\frac{3\gamma}{2}$&  $\begin{array}{c} \text{nonhyperbolic if}\;
\gamma=0\; \text{or}\;
\gamma=2\;\text{or}\;  \gamma=\frac{2}{3}\\[0.1cm]
\text{saddle, otherwise.}\\[0.1cm]\end{array}$ \\\hline
$P_3^-$& $- 6,- 3+\sqrt{\frac{3}{2}}\lambda,- 3(2-\gamma), -2$& $\begin{array}{c} \text{nonhyperbolic if}\; \gamma=2\; \text{or}\;  \lambda=\sqrt{6};\\[0.1cm]
\text{stable (node) if}\; \lambda<\sqrt{6}\; \text{and}\; \gamma\neq 2;\\[0.1cm]
\text{saddle, otherwise.}\\[0.1cm]\end{array}$ \\\hline
$P_4^-$& $-\lambda^2,\frac{1}{2}(6-\lambda^2),-\lambda^2+3\gamma, 1-\frac{\lambda^2}{2}$& $\begin{array}{c} \text{nonhyperbolic if}\; \lambda=0\; \text{or}\;  \lambda^2=3\gamma;\\[0.1cm]
\text{saddle, otherwise.}\\[0.1cm]\end{array}$ \\\hline
$P_5^-$& $-3\gamma,\frac{3}{4} \left(2-\gamma
\pm\frac{1}{\lambda}\sqrt{\Delta}\right), 1-\frac{3\gamma}{2}$&
$\begin{array}{c} \text{nonhyperbolic if}\; \gamma=0\; \text{or}\;
\lambda^2=3\gamma\;
\text{or}\;  \gamma=\frac{2}{3};\\[0.1cm]
\text{saddle,
otherwise.}\\[0.1cm]\end{array}$ \\\hline
$P_6^-$& $-2,2,-1, -2+3\gamma$& $\begin{array}{c} \text{nonhyperbolic if}\; \gamma=\frac{2}{3};\\[0.1cm]
\text{saddle, otherwise.}\\[0.1cm]\end{array}$ \\\hline
$P_7^-$& $-2,1\pm\sqrt{\frac{8}{\lambda^2}-3}, -2+3\gamma$&
$\begin{array}{c} \text{nonhyperbolic if}\; \lambda^2=2\;
\text{or}\;  \gamma=\frac{2}{3};\\[0.1cm]
\text{unstable, otherwise.}\\[0.1cm]\end{array}$
\\[0.3cm]
\hline \hline
\end{tabular}\label{negtab2}
\end{center}
\end{table*}

In table \ref{negtable9}, where we present a summary of attractors
(both past and future) for the quinstant model with $k=-1.$

\begin{table*}[ht!]\caption{Summary of attractors for for the quinstant model with $k=-1$ (system (\ref{negEqx}-\ref{negEqH})).}\label{negtable9}
\begin{tabular}{@{\hspace{4pt}}c|@{\hspace{10pt}}c@{\hspace{10pt}}c@{\hspace{2pt}}}
\hline
\hline\\[-0.3cm]
Restrictions &  Past attractor    & Future attractor \\[0.1cm]
\hline\\[-0.3cm]
${\cal H}=-1$ & $\begin{array}{cc}
P_4^- & \text{if}\;0<\gamma<2,\, \lambda^2<3\gamma\; \text{or}\\[0.2cm]
\quad & \text{if}\;\frac{2}{3}<\gamma\leq 2,\, \lambda^2<2\\[0.2cm]
P_5^- & \;0<\gamma<\frac{2}{3},\, \lambda^2>3\gamma\\[0.2cm]
P_7^- & \text{if}\;\frac{2}{3}<\gamma\leq 2,\, \lambda^2>2\\[0.2cm]
\end{array}$   & $\begin{array}{cc}
P_3^-& \text{if}\;\lambda<\sqrt{6},\,\gamma\neq 2\\[0.2cm]
P_1^- & \text{if}\;\lambda>-\sqrt{6},\,\gamma\neq 2\\[0.2cm]
\end{array}$  \\[0.2cm]\hline\\[0.1cm]
$-1<{\cal H}<1$ & $\begin{array}{cc}
P_3^+ & \text{if}\;\lambda>-\sqrt{6},\,\gamma\neq 2\\[0.2cm]
P_1^+& \text{if}\;\lambda<\sqrt{6},\,\gamma\neq 2\\[0.2cm]
\end{array}$  & as above \\[0.2cm]\hline\\[0.1cm]
${\cal H}=1$ & as above  & $\begin{array}{cc}
P_4^+ & \text{if}\;0<\gamma<2,\, \lambda^2<3\gamma\; \text{or}\\[0.2cm]
\quad & \text{if}\;\frac{2}{3}<\gamma\leq 2,\, \lambda^2<2\\[0.2cm]
P_5^+ & \;0<\gamma<\frac{2}{3},\, \lambda^2>3\gamma\\[0.2cm]
P_7^+ & \text{if}\;\frac{2}{3}<\gamma\leq 2,\, \lambda^2>2\\[0.2cm]
\end{array}$ \\[0.4cm]
\hline \hline
\end{tabular}
\end{table*}

\begin{figure}[!]
\begin{center}
\hspace{0.4cm} \put(178,165){${P_1^-}$} \put(115,133){${P_2^+}$}
\put(65,90){${P_3^+}$} \put(170,80){${P_4^+}$}
\put(115,213){${P_6^+}$} \put(40,63){${x}$} \put(235,103){${y}$}
\put(115,240){${\cal H}$}
\mbox{\epsfig{figure=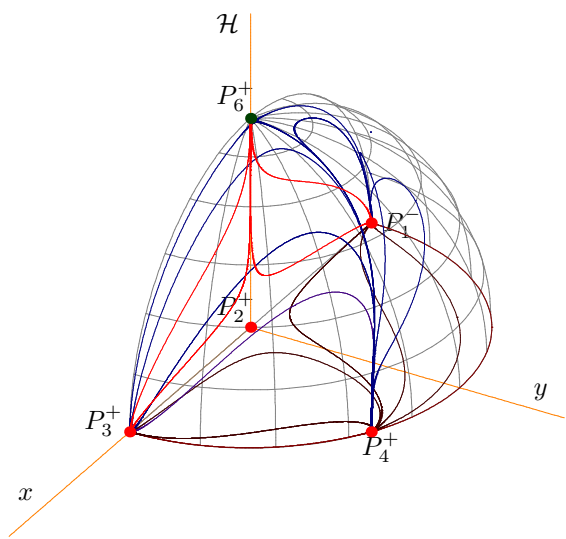,width=9cm,angle=0}} \caption{(Color
online) {\it Some orbits of (\ref{negEqx}-\ref{negEqH}) in the
invariant set ${\cal H}=1$ for the values
$\lambda=-\sqrt{\frac{3}{2}}$ and $\gamma=1.$ }}\label{Phase4}
\end{center}
\end{figure}

\begin{figure}[!]
\begin{center}
\hspace{0.4cm} \put(178,165){${P_1^-}$} \put(115,133){${P_2^-}$}
\put(65,90){${P_3^-}$} \put(223,125){${P_4^-}$}
\put(115,213){${P_6^-}$} \put(40,63){${x}$} \put(235,103){${y}$}
\put(115,240){${\cal H}$}
\mbox{\epsfig{figure=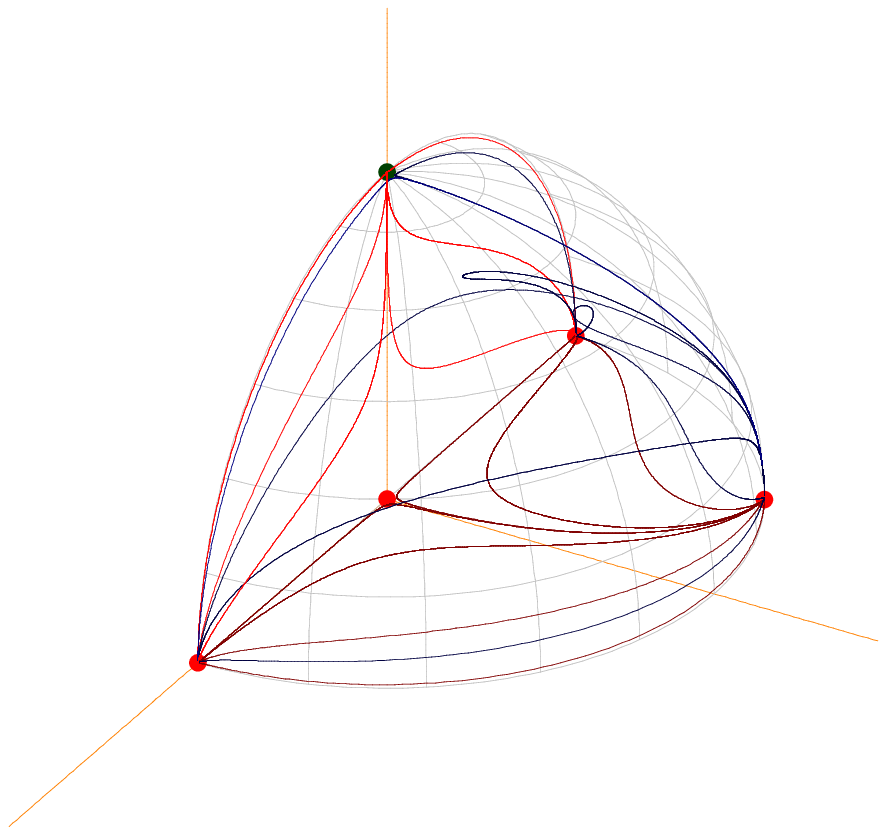,width=9cm,angle=0}} \caption{(Color
online) {\it Some orbits of (\ref{negEqx}-\ref{negEqH}) in the
invariant set ${\cal H}=-1$ for the values
$\lambda=-\sqrt{\frac{3}{2}}$ and $\gamma=1.$}}\label{Phase5}
\end{center}
\end{figure}

In figure \ref{Phase4} are displayed typical orbits of the system
(\ref{negEqx}-\ref{negEqH}) in the invariant set ${\cal H}=1$ for
the values $\lambda=-\sqrt{\frac{3}{2}}$ and $\gamma=1.$ The
critical points $P_1^+$ and $P_3^+$ are the past asymptotic
attractors, $P_1^+$ having a stronger unstable manifold tangent to
the $y$-axis. $P_4^+,$ acts as a local attractor in the invariant
set ${\cal H}=1.$ The Milne's universe ($P_6^+$) is the local
future attractor in the invariant set $y=0.$ In figure
\ref{Phase5} are drawn some orbits in the invariant set ${\cal
H=-1}$ for the same choice. The critical points $P_1^-$ and
$P_3^-$  are the future asymptotic attractors, $P_3^-$ having a
stronger attracting manifold tangent to the $y$-axis. $P_4^-$ acts
as the local source for the invariant set $z=0$ (they are,
however, saddle points for the 4D dynamics). The Milne's universe
($P_6^-$) is the local past attractor in the invariant set $y=0.$
This is the better we can do numerically since the phase space is
actually 4D.

\subsubsection{Bifurcations.}

The critical points $\left(P_4^+,\, P_4^-\right)$ reduce to
$\left(P_1^+,\, P_3^-\right)$ as $\lambda\rightarrow
(\sqrt{6})^-.$ The critical points $\left(P_4^+,\, P_4^-\right)$
reduce to $\left(P_3^+,\, P_1^-\right)$ as $\lambda\rightarrow
(-\sqrt{6})^+.$ The critical points $P_5^\pm$ reduce to $P_2^\pm$
as $\gamma\rightarrow 0^+.$ On the other hand, $P_5^\pm$ reduce to
$P_4^\pm$ as $\lambda\rightarrow (\sqrt{3\gamma})^+$ or
$\lambda\rightarrow (-\sqrt{3\gamma})^-.$ The critical points
$P_7^\pm,$ $P_5^\pm$ and $P_4^\pm$ coincide as $\gamma\rightarrow
(\frac{2}{3})^+$ and $\lambda \rightarrow (\sqrt{2})^+$
simultaneously. For these values of the parameters a bifurcation
arises.

\subsubsection{Typical behavior.}

Once the attractors have been identified one can give a
quantitative description of the physical behavior of a typical
negatively curved quinstant cosmology.

For example, for ever expanding cosmologies with $H>0,
H\rightarrow +\infty, {\cal H}=1$, i.e., the standard expanding
cosmology near the big-bang, a typical model behaves like a
massless scalar field (kinetic dominated energy density)
represented by $P_3^+$ or $P_1^+$ provided $\lambda>-\sqrt{6},\,
\gamma\neq 2$ or $\lambda<\sqrt{6},\, \gamma\neq 2,$ respectively.
This types of solutions might coexist in the same phase space. The
late time dynamics in ${\cal H}=1$ is given by either a scalar
field dominated solution ($\Omega_\phi\rightarrow 1$) represented
by $P_4^+$  provided $0<\gamma<2,\lambda^2<3\gamma$ or
$\frac{2}{3}<\gamma\leq 2,\lambda^2<2;$ or by a scaling solution
($\Omega_m/\Omega_\phi=O(1)$) represented by $P_5^+$ for
$0<\gamma<\frac{2}{3},\lambda^2>2;$ or by a curvature scaling
solution represented by $P_7^+$ provided $\frac{2}{3}<\gamma\leq
2,\lambda^2>2.$ For finite values of $H$, i.e., $-1<{\cal H}<1,$
the early time dynamics is the same as in the previous case but
there are subtle differences with respect the late time dynamics.
In fact, in the invariant set $-1<{\cal H}<1$ the future
attractors are $P_3^-$ or $P_1^-$ depending if
$\lambda<\sqrt{6},\, \gamma\neq 2$ or $\lambda>-\sqrt{6},\,
\gamma\neq 2.$ If $|\lambda|<\sqrt{6}$ the system is bistable.
Such solutions represent contracting stiff-fluid cosmologies. This
means that a typical quinstant negatively curved cosmologies
allows the collapse of matter when the time evolves. For
contracting cosmologies with ($H<0, H\rightarrow\-\infty, {\cal
H}=-1$), i.e., the standard contracting model near the initial
singularity, the late time dynamics is the same as int the
previously described case, i.e., the collapse. However, there are
subtle differences concerning the early time dynamics. The late
time dynamics in ${\cal H}=-1$ is given by either a scalar field
dominated solution ($\Omega_\phi\rightarrow 1$) represented by
$P_4^-$  provided $0<\gamma<2,\lambda^2<3\gamma$ or
$\frac{2}{3}<\gamma\leq 2,\lambda^2<2;$ or by a scaling solution
($\Omega_m/\Omega_\phi=O(1)$) represented by $P_5^-$ for
$0<\gamma<\frac{2}{3},\lambda^2>2;$ or by a curvature scaling
solution represented by $P_7^-$ provided $\frac{2}{3}<\gamma\leq
2,\lambda^2>2.$

\subsection{Quinstant cosmology with positive
curvature}\label{quinstantpos}

\subsubsection{Normalization, state space, and dynamical system.}

Let us consider the normalization factor
$\hat{D}=\sqrt{H^2-\Lambda/3+\frac{1}{a^2}}>0$ and the time
variable $d\hat{\tau}=\hat{D} dt,$ and the phase space variables:

\begin{equation}
\hat{x}=\frac{\dot\phi}{\sqrt{6} \hat{D}},\;
\hat{y}=\frac{\sqrt{V(\phi)}}{\sqrt{3} \hat{D}},\; \hat{
H}=\frac{H}{\hat{D}},\hat{z}=\frac{1}{a \hat{D}}.\label{posVars}
\end{equation}
The former variables lies in the compact phase space
$$\left\{(x,y,z): x^2+y^2\leq 1,\; y\geq 0,\; -1\leq \hat{H}
\leq 1,0\leq z\leq 1\right\}.$$ The variables
$\hat{x},\,\hat{y},\,\hat{z},$ and $\hat{H}$ satisfy the ASODE
(the prime denoting derivative with respect to $\hat{\tau}$):

\begin{eqnarray}
\hat{x}'&=&\frac{3}{2} \hat{x} {\hat{H}}
\left(\left(2-\gamma\right)\left(1-\hat{x}^2\right)-\gamma
\hat{y}^2 \right)
-\frac{\sqrt{6}}{2}\lambda \hat{y}^2,\label{posEqx}\\
\hat{y}'&=&\frac{3}{2} \hat{y}
{\hat{H}}\left(\left(2-\gamma\right)\hat{x}^2+\gamma\left(1-\hat{y}^2\right)\right)
+\frac{\sqrt{6}}{2}\lambda\hat{x}\hat{y},\label{posEqy}\\
\hat{z}'&=&\frac{1}{2}\hat{z}
{\hat{H}}\left(3\left(2-\gamma\right)\hat{x}^2+3\gamma\left(1-\hat{y}^2\right)
-2\right)\label{posEqz}\\
\hat{H}'&=&-\frac{3}{2} \left({\hat{H}}^2-1\right) \left((\gamma
-2) \hat{x}^2+\left(\hat{y}^2-1\right)
\gamma\right)+\hat{z}^2\label{posEqH}
\end{eqnarray}

As before, we will re-express the cosmological magnitudes of
interest in terms of the normalized variables.

The deceleration parameter is explicitly
\begin{equation}
q\equiv-\ddot a a/\dot a^2=-1+\frac{3}{2}\left[\frac{\hat{x}^2
\left(2-\gamma\right)+ \left(1-\hat{y}^2\right)
\gamma}{\hat{H}^2}\right]-\frac{\hat{z}^2}{\hat{H}^2};\end{equation}
the fractional energy density of the scalar field and curvature
are given respectively by

\begin{equation}
\Omega_\phi=\frac{\hat{x}^2+\hat{y}^2}{\hat{H}^2};
\Omega_k=\frac{\hat{z}^2}{\hat{H}^2}
\end{equation}
and the 'effective' EoS parameter is given by

\begin{equation}
\omega_{eff}\equiv\frac{P_{tot}}{\rho_{tot}}\equiv\frac{\frac{1}{2}
\dot\phi^2-V(\phi )+(\gamma -1) \rho_{\rm M} -\Lambda
}{\frac{1}{2} \dot\phi ^2+V(\phi )+\rho_{\rm M} +\Lambda
}=-1+\frac{\left(2-\gamma\right)
\tilde{x}^2+\left(1-\tilde{y}^2\right)
\gamma}{\hat{H}^2+\hat{z}^2}.
\end{equation}

\subsubsection{Form invariance under coordinate transformations.}

The system (\ref{posEqx}-\ref{posEqH}) is form invariant under the
coordinate transformation and time reversal

\begin{equation}\left(\hat{\tau},\,\hat{x},\,\hat{y},\,\hat{z},\,\hat{H}\right)\rightarrow
\left(-\hat{\tau},\,-\hat{x},\,\hat{y},\,\hat{z},\,-\hat{H}\right).\label{posdiscrete4}\end{equation}

Thus, it is sufficient to discuss the behavior in one part of the
phase space, the dynamics in the other part being obtained via the
transformation (\ref{posdiscrete4}). Observe that equations
(\ref{posEqx}-\ref{posEqH}) are form invariant under the
coordinate transformation $\hat{y}\rightarrow -\hat{y}$ and
$\hat{z}\rightarrow -\hat{z}.$ Then, (\ref{posEqx}-\ref{posEqH})
is form invariant under they composition with
(\ref{posdiscrete4}).

There are two obvious invariant sets under the flow of
(\ref{posEqx}-\ref{posEqH}), they are $\hat{y}=0,$ $\hat{z}=0.$

From equation (\ref{posEqH}) we can immediately see that the
surfaces $\hat{H}=\pm 1$ are not invariant provided $z\neq 0.$ In
fact, the surfaces $\hat{H}=\pm 1$ act as membranes (that can be
crossed). This follows from the fact that $H'|_{H=\pm 1}=z^2>0$
for $z\neq 0.$ Observe that if initially $z>0,$ then, from
equation (\ref{posEqz}), follows that the sign of $z$ is
invariant. Only if $z=0,$ the surfaces $\hat{H}=\pm 1$ could be
invariant.

\subsubsection{Monotonic functions}

Let be defined in the invariant set
$$\left\{(\hat{x},\hat{y},\hat{z},\hat{H}): \hat{x}^2+\hat{y}^2< 1,\; \hat{y}> 0,\; 0<z<1,\; -1<\hat{H}<
1\right\},$$ the function

\begin{equation}
Z=\frac{z^4}{\left(1-x^2-y^2\right)^2},\;
Z'=-2\left(2-3\gamma\right)\hat{H}Z\label{posZ}
\end{equation} This function is monotonic in the regions $\hat{H} < 0$ and $\hat{H}>0$ for $\gamma\neq \frac{2}{3}.$

The existence of this monotonic function allows to state that
there can be no periodic orbits or recurrent orbits in the
interior of the phase space. Furthermore, it is possible to obtain
global results. From the expression $Z$ we can immediately see
that asymptotically $z\rightarrow 0$ or $x^2+y^2\rightarrow 1.$

\subsubsection{Local analysis of critical points.}

\begin{table*}[ht]\caption[crit]{Location and existence conditions of the critical points of the system
(\ref{posEqx}-\ref{posEqz})}
\begin{center}
\begin{tabular}{@{\hspace{4pt}}c@{\hspace{14pt}}c@{\hspace{14pt}}c@{\hspace{2pt}}}
\hline
\hline\\[-0.3cm]
Label & Coordinates: $(\hat{x},\hat{y}, \hat{z},\hat{H})$& Existence\\[0.1cm]
\hline\\[-0.2cm]
$\hat{P}_1^\pm$& $(-1,0,0,\pm 1)$& All $\lambda$  \\\hline\\
$\hat{P}_2^\pm$& $(0,0,0,\pm 1)$& All $\lambda$  \\\hline\\
$\hat{P}_3^\pm$ & $(1,0,0,\pm 1)$& All $\lambda$   \\\hline\\
$\hat{P}_4^\pm$&
$\left(\mp\frac{\lambda}{\sqrt{6}},\sqrt{1-\frac{\lambda^2}{6}},0,\pm 1\right)$& $-\sqrt{6}< \lambda<\sqrt{6}$ \\\hline\\
$\hat{P}_5^\pm$&
$\left(\mp\sqrt{\frac{3}{2}}\frac{\gamma}{\lambda},\sqrt{\frac{3}{2}}
\sqrt{\frac{(2-\gamma)\gamma}{\lambda^2}},0,\pm 1\right)$ &
$\begin{array}{c}
\gamma=0,\;\lambda\neq 0\\[0.2cm]
0<\gamma\leq 2,\; |\lambda|  \geq \sqrt{3\gamma}\\[0.2cm]
\end{array} $\\\hline\\
$\hat{P}_6$& $\left(x^\star, 0,
\sqrt{\frac{3}{2}\left(\gamma+(2-\gamma){x^\star}^2\right)},0\right)$
& $
0\leq \gamma\leq \frac{2}{3},\; |x^\star|  \leq \sqrt{\frac{2-3\gamma}{3(2-\gamma)}}$ \\\hline\\
$\hat{P}_7^\pm$&
$\left(\mp\frac{1}{\sqrt{3}},\sqrt{\frac{2}{3}},\sqrt{1-\frac{\lambda^2}{2}},\pm
\frac{\lambda}{\sqrt{2}}\right)$& $-\sqrt{2}\leq \lambda\leq
\sqrt{2}$
\\[0.4cm]
\hline \hline
\end{tabular}\label{postab1}
\end{center}
\end{table*}

\begin{table*}[ht]\caption[crit2]{Eigenvalues, and dynamical character of the fixed points of the system (\ref{negEqx}-\ref{negEqH}).
We use the notation
$\Delta=(2-\gamma)(24\gamma^2+\lambda^2(2-9\gamma))$ and
$\mu=\sqrt{\frac{3}{2}\left(\left({x^\star}^2-1\right)\gamma\left(2-3\gamma\right)+24
{x^\star}^2\right)}.$}
\begin{center}
\begin{tabular}{@{\hspace{2pt}}c@{\hspace{10pt}}c@{\hspace{14pt}}c@{\hspace{2pt}}}
\hline
\hline\\[-0.3cm]
Label & Eigenvalues& Dynamical character\\[0.1cm]
\hline\\[-0.2cm]
$\hat{P}_1^-$& $- 6,- 3-\sqrt{\frac{3}{2}}\lambda,- 3(2-\gamma), -2$& $\begin{array}{c} \text{nonhyperbolic if}\; \gamma=2\; \text{or}\;  \lambda=-\sqrt{6};\\[0.1cm]
\text{stable (node) if}\; \lambda>-\sqrt{6}\; \text{and}\; \gamma\neq 2;\\[0.1cm]
\text{saddle, otherwise.}\\[0.1cm]\end{array}$\\\hline
$\hat{P}_2^-$& $\frac{3}{2}(2-\gamma),- 3\gamma,-
\frac{3\gamma}{2},1- \frac{3\gamma}{2}$&  $\begin{array}{c}
\text{nonhyperbolic if}\; \gamma=0\; \text{or}\;
\gamma=2\;\text{or}\;  \gamma=\frac{2}{3}\\[0.1cm]
\text{saddle otherwise.}\\[0.1cm]\end{array}$ \\\hline
$\hat{P}_3^-$& $- 6,- 3+\sqrt{\frac{3}{2}}\lambda,- 3(2-\gamma), -2$& $\begin{array}{c} \text{nonhyperbolic if}\; \gamma=2\; \text{or}\;  \lambda=\sqrt{6};\\[0.1cm]
\text{stable (node) if}\; \lambda<\sqrt{6}\; \text{and}\; \gamma\neq 2;\\[0.1cm]
\text{saddle, otherwise.}\\[0.1cm]\end{array}$ \\\hline
$\hat{P}_4^-$&
$-\lambda^2,\frac{1}{2}(6-\lambda^2),-\lambda^2+3\gamma,
1-\frac{\lambda^2}{2}$& $\begin{array}{c} \text{nonhyperbolic
if}\; \lambda=0\; \text{or}\;
\lambda^2=3\gamma\; \text{or}\;  \lambda^2=2 ;\\[0.1cm]
\text{saddle otherwise.}\\[0.1cm]\end{array}$ \\\hline
$\hat{P}_5^-$& $-3\gamma,\frac{3}{4} \left(2-\gamma
\pm\frac{1}{\lambda}\sqrt{\Delta}\right), 1-\frac{3\gamma}{2}$&
$\begin{array}{c} \text{nonhyperbolic if}\; \gamma=0\; \text{or}\;
\lambda^2=3\gamma\;
\text{or}\;  \gamma=\frac{2}{3};\\[0.1cm]
\text{saddle,
otherwise.}\\[0.1cm]\end{array}$ \\\hline
$\hat{P}_6$& $0, \sqrt{\frac{3}{2}}\lambda x^\star,-\mu, \mu$&
nonhyperbolic
\\\hline
$\hat{P}_7^-$& $-\sqrt{2} \lambda ,\frac{\lambda \pm\sqrt{8-3
\lambda ^2}}{\sqrt{2}},\frac{(3 \gamma-2) \lambda }{\sqrt{2}}$&
$\begin{array}{c} \text{nonhyperbolic if}\; \lambda=0\;\text{or}\;
\lambda^2=2;
\text{or}\;  \gamma=\frac{2}{3};\\[0.1cm]
\text{unstable (saddle), otherwise.}\\[0.1cm]\end{array}$
\\[0.3cm]
\hline \hline
\end{tabular}\label{postab2}
\end{center}
\end{table*}

Using (\ref{posdiscrete4}), it is possible to infer the local
stability of $\hat{P}_1^+, \hat{P}_2^+, \hat{P}_3^+, \hat{P}_4^+,
\hat{P}_5^+, \hat{P}_7^+$ from the local stability of $P_3^-,
\hat{P}_2^-, \hat{P}_1^-, \hat{P}_4^-, \hat{P}_5^-, \hat{P}_7^-.$
Thus, we will analyze only the critical points in the ``negative''
branch.

The critical points $\hat{P}_1^+$ to $\hat{P}_5^+$ have similar
properties, dynamical character as unhatted ones in section
\ref{quinstantneg} and the same physical interpretation of the
unhatted ones characterized in section \ref{Flat}. Thus, we will
not comment about its stability in detail. For completeness in
table \ref{postab2} are summarize the existence conditions and
stability properties. We will submit the interested reader to
previous sections for more details.

For the positive curvature model there are new critical points:
the curve of nonhyperbolic critical points $\hat{P}_6$ represent
the Einstein's static universe. Special critical points of this
family are those with the choice $x^\star=0$ (it exists provided
$0\leq\gamma\leq \frac{2}{3}$) and
$x^\star=\pm\sqrt{\frac{3(2-3\gamma)}{2-\gamma}}$ (it exists
provided $0\leq\gamma< \frac{2}{3}$), and the curvature-scaling
solution $\hat{P}_7^\pm.$ Concerning the stability of
curvature-scaling solutions we have that $P_7^-$ is always a
saddle point (at least one eigenvalues has negative real part, and
the others are of different sign). Its stable manifold is 1D
provided $-\sqrt{2}<\lambda<0,\, 0<\gamma\frac{2}{3};$ or 2D
provided $0<\lambda<\sqrt{2},\, \frac{2}{3}<\gamma<2$ or
$-\sqrt{2}<\lambda<0,\,\frac{2}{3}<\gamma<2;$ or 3D if
$0<\lambda<\sqrt{2},\, 0\leq\gamma<\frac{2}{3}.$

\subsubsection{Bifurcations.}

The critical points $\left(\hat{P}_4^+,\, \hat{P}_4^-\right)$
reduce to $\left(\hat{P}_1^+,\, \hat{P}_3^-\right)$ as
$\lambda\rightarrow (\sqrt{6})^-.$ The critical points
$\left(\hat{P}_4^+,\, \hat{P}_4^-\right)$ reduce to
$\left(\hat{P}_3^+,\, \hat{P}_1^-\right)$ as $\lambda\rightarrow
(-\sqrt{6})^+.$ The critical points $\hat{P}_5^\pm$ reduce to
$\hat{P}_2^\pm$ as $\gamma\rightarrow 0^+.$ On the other hand,
$\hat{P}_5^\pm$ reduce to $\hat{P}_4^\pm$ as $\lambda\rightarrow
(\sqrt{3\gamma})^+$ or $\lambda\rightarrow (-\sqrt{3\gamma})^-.$
$(\hat{P}_7^+,\hat{P}_7^-)$ reduce to $(\hat{P}_4^+,\hat{P}_4^-)$
as $\lambda\rightarrow\left(\sqrt{2}\right)^-$ and
$(\hat{P}_7^+,\hat{P}_7^-)$ reduce to $(\hat{P}_4^-,\hat{P}_4^+)$
as $\lambda\rightarrow\left(-\sqrt{2}\right)^+.$ For these values
of the parameters a bifurcation arises.

\subsubsection{Typical behavior.}

As a consequence that $\hat{H}=\pm 1$ are not invariant sets, the
determination of past and future attractors is more simpler. If
$\lambda<-\sqrt{6}, \gamma\neq 2,$ then $P_1^+$ is the past
attractor and $P_3^-$ is the future attractor. If
$-\sqrt{6}<\lambda<\sqrt{6},\,\gamma\neq 2$ the past attractors
are both $\hat{P}_3^+$ and  $\hat{P}_1^+$ and the future
attractors are  both $\hat{P}_1^-$ and $\hat{P}_3^-.$ Finally, if
$\lambda>\sqrt{6}, \gamma\neq 2$ then  $P_3^+$ is the past
attractor and $P_1^-$ is the future attractor. In any case the
Universe evolves from a stiff regime to a stiff regime by crossing
the value $\hat{H}=0,$ allowing the collapse of the Universe.

In table \ref{postable9}, where we present a summary of attractors
(both past and future) for the quinstant model with $k=1.$

\begin{table*}[ht!]\caption{Summary of attractors for for the quinstant model with $k=1$
(system (\ref{posEqx}-\ref{posEqH})) for $z>0.$}\label{postable9}
\begin{center}
\begin{tabular}{@{\hspace{4pt}}c|@{\hspace{10pt}}c@{\hspace{10pt}}c@{\hspace{2pt}}}
\hline
\hline\\[-0.3cm]
Past attractor    & Future attractor \\[0.1cm]
\hline\\[-0.3cm]
$\begin{array}{cc}
\hat{P}_3^+ & \text{if}\;\lambda>-\sqrt{6},\,\gamma\neq 2\\[0.2cm]
\hat{P}_1^+& \text{if}\;\lambda<\sqrt{6},\,\gamma\neq 2\\[0.2cm]
\end{array}$  & $\begin{array}{cc}
\hat{P}_3^-& \text{if}\;\lambda<\sqrt{6},\,\gamma\neq 2\\[0.2cm]
\hat{P}_1^- & \text{if}\;\lambda>-\sqrt{6},\,\gamma\neq 2\\[0.2cm]
\end{array}$
\\[0.4cm]
\hline \hline
\end{tabular}
\end{center}
\end{table*}

\section{Observational Test and Dynamical Systems: The Interplay}

Dynamical systems techniques by one way and Observational test by
the other are strongly enough to discriminate among the wide
variety of dark energy models nowadays under investigation. The
first one is more mathematical in character: they can be used to
select the better behaved models, with appropriate attractors in
the past and future. These tools to analyse and interpreted
results has gained a lot of attention in recent years. The
techniques are so powerful when we want to investigate the
asymptotic (and even, intermediate) behavior of models. The other
class of tools is rather physical: they can be used of
astrophysical observations to crack the degeneracy of classes of
dark energy models. In the interplay, both serve to constraint the
free parameters of the models under consideration. In our case,
quintom and quinstant dark energy, with flat and curved
geometry.\\

\underline{Quintom Dark Energy Paradigm}
\\

The model of quintom, which is mainly favored by current SNIa
only, needs to be confronted with other observations in the
framework of concordance cosmology. Since SNIa offer the only
direct detection of DE, this model is the most promising to be
distinguished from the cosmological constant and other dynamical
DE models which do not get across $-1$, by future SNIa projects on
the low redshift (for illustrations see e.g.
\cite{Huterer:2004ch}).

From the dynamical systems viewpoint we have obtained further
results in support of the previous results in
\cite{Guo:2004fq,Zhang:2005eg,Lazkoz:2006pa}. For negative
curvature models, we have devised two dynamical systems adapted to
the study of expanding ($\epsilon=\text{sign}\,{H}>0$) and
contracting ($\epsilon=\text{sign}\,{H}<0$) models. Also, we have
devised another dynamical system well suited for investigating
positive curvature models. We have characterized the critical
points of each system and interpreted the cosmological solutions
associated. By devising well defined monotonic functions we were
able to get global results for ever expanding and contracting
models (for both negative and positive curvature models). We have
reviewed the results concerning the flat case. It is known that,
for flat ever expanding models the attractor will be the matter
scaling solution \cite{Lazkoz:2006pa}. If matter scaling solutions
do not exist, the attractor will be phantom ($w<-1$) or de Sitter
($w=-1)$ like. This is a difference with respect to the results in
\cite{Guo:2004fq} and \cite{Zhang:2005eg}. It was proved there,
that the attractor solutions are de Sitter-like, unless some
trajectories cross, transiently,  the $w=-1$ boundary to become
even smaller before ending in a de Sitter phase.

The new results we survey here are as follows:

For negative-curvature ever-expanding models
($\epsilon=\text{sign}\,{H}>0$) we have obtained the existence of
scaling curvature attractors (without matter) (provided
$\delta<\frac{1}{3}$). The attractor solution will be dominated by
DE whenever its existence precludes the existence of scaling
curvature attractors. These solutions can be: phantom-like
($w<-1$), de Sitter-like ($w=-1$), or quintessence-like. This is a
difference with respect the situation in \cite{Lazkoz:2006pa}. We
must notice, however, that if we consider other values for
$\gamma,$ other than $\gamma=1,$ then the attractor of the system
can be the matter scaling solution. This is the case if
$0<\gamma<\sfrac{2}{3},\,\delta>\sfrac{\gamma}{2}.$ Under the
above conditions on the parameters DM mimics DE. For contracting
models ($\epsilon=\text{sign}\,{H}<0$), the attractor will be a
MSF solution that mimics a stiff fluid. Towards the past, the
typical situation is the reverse of the former described. For
positive curvature (closed) models, we have obtained conditions
under which there is an orbit of type ${}_+\hat{K}\rightarrow
{}_-\hat{K}.$ This represents a cosmological solution starting in
a ending towards a singularity describe a  MSF cosmology. We have
obtained, also, a flat FRW solution starting in a  big-bang in
${}_+F$ y recolapsing in a ``big-crunch'' in ${}_-F.$ We have
illustrated this results by means of numerical integrations of the
ASODE describing this cosmological model. We have obtained
conditions for the existence of global attractors. We have
offered, here, only a simplified qualitative analysis (as a
difference of the mathematical analysis in \cite{Coley:2003mj}
pages 69-73). However, our study have relevance by its own right,
and can be considered in some way as a complement of the former
since we have added a phantom field in the dynamics. We must to
restate, however, that our analysis is not as detailed as in that
reference. But its is suffice to illustrated our goals. The
qualitative analysis in multi- scalar field (coventional)
cosmologies with exponential potentials (in the context of
assisted inflation) was done in the same reference, section VII,
and in \cite{vdhoogen,Coley:1999mj}, particularly for two fields.
They do not consider phantom field as we do here. Our monotonic
functions were able to discard the existence of periodic orbits,
homoclinic orbits, or recurrent orbits.\\

\underline{Quinstant Dark Energy Paradigm}
\\

From the stability analysis of all
studied models of (exponential) quinstant dark energy
(\cite{RolLambda}, \cite{Cardone2008}), the typical
behavior, irrespective the curvature choice, is the evolution from
an stiff-regime near the past, to a stiff-regime in the far
future. Besides, from our dynamical analysis it seems that for
negatively curvature and flat models, the model shows the
divergence of the Hubble parameter ($H$) in the asymptotic
regimes. There are other cosmological models of composite dark
energy having stiff-matter domination as an attractor in the past,
but usually they would not be global attractors, but local. This
is the case of several models of quintom dark energy. From the
structure formation we see that QDE makes reasonable predictions
for the formation of linear large scale structure of the Universe.
It reproduces reasonably well the non-linear structures from today
up to redshifts a bit larger than one, but fails to reproduce the
perturbations in the non-linear regime for redshifts a bit larger
than one. This models are dynamically equivalent models of $f(R)$
modified gravity. It would be interesting to study how these
$f(R)$ models behave concerning structure formation, and then we
would have a better understanding on how these observations would
crack the degeneracy dark energy-$f(R)$ modified gravity.

In summary, the new results concerning quinstant dark energy are
as follows:

For the standard flat expanding cosmology near the big-bang, a
typical model behaves like a massless scalar field (kinetic
dominated energy density) and the late time dynamics is given by
either a scalar field dominated solution ($\Omega_\phi\rightarrow
1$) or by a scaling solution ($\Omega_m/\Omega_\phi=O(1)$)
represented by $P_5^+$ provided $\lambda^2<3\gamma.$ This is the
standard behavior for quintessence models (without $\Lambda$). For
finite values of $H$, the early time dynamics is the same as in
the previous case but there are subtle differences with respect
the late time dynamics. In fact, in this invariant set the future
attractors are stiff-like (contracting) solutions with
$H\rightarrow \pm \infty.$ This means that a typical quinstant
cosmologies allows the collapse of matter when the time evolves.
For the standard contracting model near the initial singularity,
the late time and early time dynamics is the reverse of the
previously described.

The behavior of a typical negatively curved quinstant model is
similar to the flat situation, but not the same. The differences
is in that in the limit $H\rightarrow\infty$ the late time
dynamics given by either a scalar field dominated solution
($\Omega_\phi\rightarrow 1$) provided
$0<\gamma<2,\lambda^2<3\gamma$ or $\frac{2}{3}<\gamma\leq
2,\lambda^2<2;$ or by a scaling solution
($\Omega_m/\Omega_\phi=O(1)$) for
$0<\gamma<\frac{2}{3},\lambda^2>2;$ or by a curvature scaling
solution provided $\frac{2}{3}<\gamma\leq 2,\lambda^2>2.$ For
finite values of $H$, the future attractors are stiff-like
solutions. For positive curvature models the Universe evolves from
a stiff regime to a stiff regime by crossing the value
$\hat{H}=0,$ allowing the collapse of the Universe. Thus, from the
dynamical view point there are not significant differences between
quinstant and quintom dark energy paradigms.

Our opinion is that any dark energy model which presents a
stiff-like equation of state in the past, during a long period of
time, will predict abrupt peaks of structure formation, which
would be the result of enhanced accretion of the forming
structures, both because of gravitational and viscous forces.

\end{document}